%% file: Kling_FEL.tex
\documentclass[twocolumn,amsmath,amssymb,aps,pra,superscriptaddress,showpacs,10pt]{revtex4-1}

\usepackage{graphicx}
\usepackage{grffile} 
\usepackage{array}
\usepackage{bbm}
\usepackage{color}
\usepackage[utf8]{inputenc}
\usepackage[OT1]{fontenc}

\usepackage{units} 
\usepackage{upgreek} 

\usepackage{dsfont}
\usepackage{hyperref}

\usepackage[showonlyrefs]{mathtools}		
\mathtoolsset{showonlyrefs=true}

\renewcommand{\L}{\text{L}}
\newcommand{\W}{\text{W}}

 \newcommand{\bra}[1]{\left\langle{#1}\right|}
 \newcommand{\ket}[1]{\left|{#1}\right\rangle}
 \newcommand{\braket}[2]{\langle{#1}|{#2}\rangle}

\newcommand{\e}[1]{\operatorname{e}^{#1}}

\newcommand{\I}{i}

\newcommand{\D}{\text{d}}

\def\ba#1\ea{\begin{align}#1\end{align}}																
\newcommand{\del}{\Delta}
\newcommand{\taub}{\tau}

\usepackage{braket}

\usepackage{type1cm}
\usepackage{eso-pic}
\usepackage{xcolor}

\usepackage{tikz}                       
\usepackage{pgfplots}                   
\usetikzlibrary{spy}
\usepgfplotslibrary{external}

\usetikzlibrary{shapes,arrows,shadows,3d,calc}
\usetikzlibrary{decorations.pathmorphing,decorations.pathreplacing,decorations.markings,trees}
\usetikzlibrary{pgfplots.groupplots}

\pgfplotsset{%
  every axis legend/.append style={%
    cells={anchor=west},
    at={(0.96,0.04)},
    anchor=south east,
    font=\scriptsize
  },
  every axis/.append style={%
    yticklabel style={/pgf/number format/fixed zerofill, /pgf/number format/precision=2}
  },
  width= 0.45\textwidth, height=5cm, xmajorgrids=false, xminorgrids=false, minor x tick num=1
}

\usepackage[hang]{subfigure} 

\pgfplotsset{compat=1.8}

\usepackage{blindtext} 
\usepackage{booktabs}

\begin{document}
\title{A high-gain Quantum free-electron laser: emergence \& exponential gain}
\author{Peter Kling}
\affiliation{Institut für Quantenphysik and Center for Integrated Quantum Science and Technology $\left(\text{IQ}^{\text{ST}}\right)$, Universität Ulm, Albert-Einstein-Allee 11, D-89081, Germany}      \affiliation{Helmholtz-Zentrum Dresden-Rossendorf eV, D-01328 Dresden, Germany}

\author{Enno Giese}
\affiliation{Institut für Quantenphysik and Center for Integrated Quantum Science and Technology $\left(\text{IQ}^{\text{ST}}\right)$, Universität Ulm, Albert-Einstein-Allee 11, D-89081, Germany} 

\author{C. Moritz Carmesin}
\affiliation{Helmholtz-Zentrum Dresden-Rossendorf eV, D-01328 Dresden, Germany}
\affiliation{Institut für Quantenphysik and Center for Integrated Quantum Science and Technology $\left(\text{IQ}^{\text{ST}}\right)$, Universität Ulm, Albert-Einstein-Allee 11, D-89081, Germany} 

\author{Roland Sauerbrey}
\affiliation{Helmholtz-Zentrum Dresden-Rossendorf eV, D-01328 Dresden, Germany}

\author{Wolfgang P. Schleich}
 \affiliation{Institut für Quantenphysik and Center for Integrated Quantum Science and Technology $\left(\text{IQ}^{\text{ST}}\right)$, Universität Ulm, Albert-Einstein-Allee 11, D-89081, Germany} 
\affiliation{Hagler Institute for Advanced Study at Texas A\&M University, Texas A\&M AgriLife Research, Institute for Quantum Science and Engineering (IQSE) and Department of Physics and Astronomy, Texas A\&M University, College Station, Texas 77843, USA }

%

\begin{abstract}
We derive an effective Dicke model in momentum space to describe collective effects in the quantum regime of a free-electron laser (FEL). The resulting exponential gain from a single passage of electrons allows the operation of a Quantum FEL in the high-gain mode and avoids the experimental challenges of an X-ray FEL oscillator. Moreover, we study the intensity fluctuations of the emitted radiation which turn out to be super-Poissonian.   
\end{abstract}

\maketitle

\input{Introduction.tex}

\input{Many_electron_model.tex}

\input{Exponential_gain.tex}

\input{Conclusions.tex}

\begin{acknowledgements}
   We thank P. Anisimov, W. Becker, M. Bussmann, A. Debus, R. Endrich, A. Gover, Y. Pan, P. Preiss, K. Steiniger, and S. Varr{\'o} for many fruitful discussions.  
   W.\,P.\,S. is grateful to Texas A\&M University for a Faculty Fellowship at the 
Hagler Institute of Advanced Study at Texas A\&M University, and Texas A\&M AgriLife 
Research for the support of his work. The Research of $\mathrm{IQ}^{\mathrm{ST}}$ is 
financially supported by the Ministry of Science, Research and Arts Baden-Württemberg.
\end{acknowledgements}

\begin{appendix}
\input{Transformation_of_Hamiltonian_in_terms_of_jump_operator.tex}
\input{Canonical_averaging.tex}

\end{appendix}

\bibliography{bib}

\end{document}

%% file: Introduction.tex
\section{Introduction}
\label{sec:Introduction}

The current trend of decreasing free-electron laser (FEL) wavelengths down to the X-ray regime~\cite{huang,pellegrini_rmp,dattoli2018} leads inevitably to a limit, where quantum effects emerge and at some point determine the dynamics and the properties of the FEL.
Since there exist no high-quality cavities for exactly this part of the spectrum, such a Quantum FEL~\cite{gover,schroeder,boni06} necessarily has to be operated in the high-gain regime, where a many-electron theory becomes mandatory~\cite{bnp}.



As a consequence of this trend, many theoretical models towards the Quantum FEL were developed in recent years~\cite{schroeder,av02,*av07,boni06,pio_prl,pio,gaiba,serbeto09,eli1,*eli2,NJP2015,anisimov2017,boni2017,
brown2017,gover2017,applb,robb2018,fares2018,boni_epl,*fares_wigner}. Besides investigating the emergence of quantum features, it was studied how these effects alter the radiation properties of an FEL. For example, it was predicted in Ref.~\cite{boni06} that a Quantum FEL operating  in the self-amplified spontaneous emission (SASE) mode emits light with a higher degree of temporal coherence and with a narrower spectrum when compared to its classical counterpart.

In contrast to a SASE FEL, the low-gain regime of operation has a well defined cavity mode. In this case, the amplification of the laser field is comparably small since the electrons travel through a relatively short undulator~\cite{schmueser}. Similar to an ordinary laser~\cite{scullylamb} the cavity serves the purpose to store the laser field that is amplified during many passages of electrons. This necessity for a cavity inhibits the operation of such an `FEL oscillator' in the X-ray regime, where up to date no high-quality mirrors exist. 

Existing models~\cite{schroeder,boni06} of the Quantum FEL focus mainly on the opposite mode of operation,  the high-gain FEL, where the radiation grows exponentially along the length of a long wiggler and a single passage of electrons is sufficient to obtain a large laser intensity~\cite{schmueser} -- without any cavity. In this sense, such a device is an `amplifier' rather than a `laser', even though the latter term is commonly used in literature. 

At the beginning of the new century,  FEL physics experienced an immense leap forward with the first lasing of X-ray FELs~\cite{emma}. The associated decrease in wavelength leads to an increased quantum mechanical recoil of the electrons when they scatter from the fields. 
Hence, an experimental realization of the quantum regime, where this recoil dominates, is within reach, apart from  today still challenging constraints on the electron beam and the undulator~\cite{debus2018}. 

In this article we adjust the elementary approach of Ref.~\cite{NJP2015} to the many-electron case by introducing suitable collective operators. We observe an exponential growth of the laser intensity with a possible start-up from vacuum and study the photon statistics of the emitted radiation.

This article is organized as follows: we begin in Sec.~\ref{sec:Many_electron_model} by discussing the implications of a many-electron theory and by extending our previous model for the Quantum FEL to the collective case. In this context, we deduce the conditions for the FEL dynamics to reduce to the two-level behavior dictated by the Dicke Hamiltonian~\cite{dicke}. In Sec.~\ref{sec:Exponential_gain}
we then solve the resulting equations of motion in the short-time limit and observe an exponential gain of the laser intensity as well as a super-Poissonain behavior of the corresponding fluctuations. Moreover, we make use of the asymptotic method of canonical averaging~\cite{bogoliubov,higher} to calculate higher-order corrections to the deep quantum regime and we connect our results to the existing literature on the Quantum FEL~\cite{boni06}. 
Finally, we summarize our main results and conclude in Sec.~\ref{sec:Conclusions}.

To keep this article self-contained we add App.~\ref{app:Transformation_of_Hamiltonian_in_terms_of_jump_operator}, where we express the FEL Hamiltionian in terms of momentum jump operators and perform a transformation into a rotating frame. In App.~\ref{app:Canonical_averaging} we recall the method of canonical averaging~\cite{higher} and apply it on the FEL. 

%% file: Many_electron_model.tex
\section{Many-electron model}
\label{sec:Many_electron_model}

We begin our studies of the high-gain Quantum FEL by presenting the many-particle FEL Hamiltonian and investigating the emergence of collective effects due to the action of this Hamiltonian on quantum states. Moreover, we introduce collective momentum jump operators  and derive the two-level behavior in the quantum regime.

\subsection{One electron vs. many electrons}

In the following we discuss the fundamental differences between  a collective model of the FEL and our previous single-electron approach~\citep{NJP2015}. 

\subsubsection{Hamiltonian}

The many-electron Hamiltonian describing the dynamics of an  FEL reads~\cite{becker83,*becker87}
\begin{equation}\label{eq:H_many}
\hat{H}=\sum\limits_{j=1}^N\frac{\hat{p}_j^2}{2m}
+\hbar g \left(\hat{a}_\L \sum\limits_{j=1}^N \e{\I 2k\hat{z}_j}
+\hat{a}_\L^\dagger \sum\limits_{j=1}^N \e{-\I 2k\hat{z}_j}\right),
\end{equation}
where we sum over  the positions $\hat{z}_j$ and conjugate momenta $\hat{p}_j$ of $N$ electrons with mass $m$  in the co-moving Bambini--Renieri frame~\cite{bambi,*brs,frame}. This frame of reference is constructed such that the wave numbers of the laser field  and of the wiggler field, $k_\L$ and $k_\W$, respectively, coincide, that is $k_\L=k_\W\equiv k$.

While we assume that the wiggler field is classical and fixed due to its high intensity, we describe the laser field by the photon annihilation and creation operator,  $\hat{a}_\L$ and $\hat{a}_\L^\dagger$, respectively. The commutation relations of the involved operators are given by $[\hat {z}_j,\hat{p}_j]=\I\hbar$ for the electron variables~\footnote{We consider the electrons as distinguishable particles and thus treat them in first quantization. Hereby, we follow the lines of Ref.~\cite{bonifacio-basis} in which it was argued that for realistic electron beams the number $N$ of electrons in a bunch is much smaller than the phase space volume of the bunch in units of $\hbar$. Therefore, we can neglect the fermionic nature of the electrons.}, with $\hbar$ denoting the reduced Planck constant, and by $[\hat{a}_\L,\hat{a}_\L^\dagger]=1$ for the laser-field operators.

We note that the constant
\ba 
g\equiv \frac{e^2\mathcal{A}_\L \tilde{\mathcal{A}}_\W}{\hbar m}
\ea
that couples the dynamics of the electrons to the fields depends on the product of the vacuum amplitude $\mathcal{A}_\L$ of the vector potential for the laser field and the classical amplitude $\tilde{\mathcal{A}}_\W$ of the vector potential for the wiggler field.

By  considering the Bambini--Renieri frame we follow the lines of a large part of the literature~\cite{ciocci,*dattoli_hb4,becker83,*becker87,becker82,*becker_pstat,*becker_lw,orszag_lw} on the quantum theory of the FEL. We note, however, that such a quantum theory can be also formulated in the laboratory  frame~\cite{gover,madey,becker1979,*becker1980,*becker80,*becker88,mciver79quantum,*fedorov_multi,
banacloche,*banacloche_pstat,gover_lw,kurizki,applb}.

\subsubsection{Low and high gain}

If we assume that the relative change of the laser intensity during one passage of a bunch with $N$ electrons is much smaller than unity, the equations of motion
for the electrons decouple from each other. In this case, we only solve the dynamics of one electron interacting with the fields: this limit is the low-gain regime of the FEL~\cite{schmueser}, which is a suitable description of an FEL oscillator, where the high-intensity field in the cavity does not vary much from one passage of electrons to the next one.

In contrast, if the initial field is very small or even starts from vacuum  we certainly cannot consider the \textit{relative} change of the laser intensity as a small quantity. In this situation, we have to solve the full many-particle Hamiltonian in Eq.~\eqref{eq:H_many}.

\begin {figure}
\centering
\includegraphics[width=0.25 \textwidth]{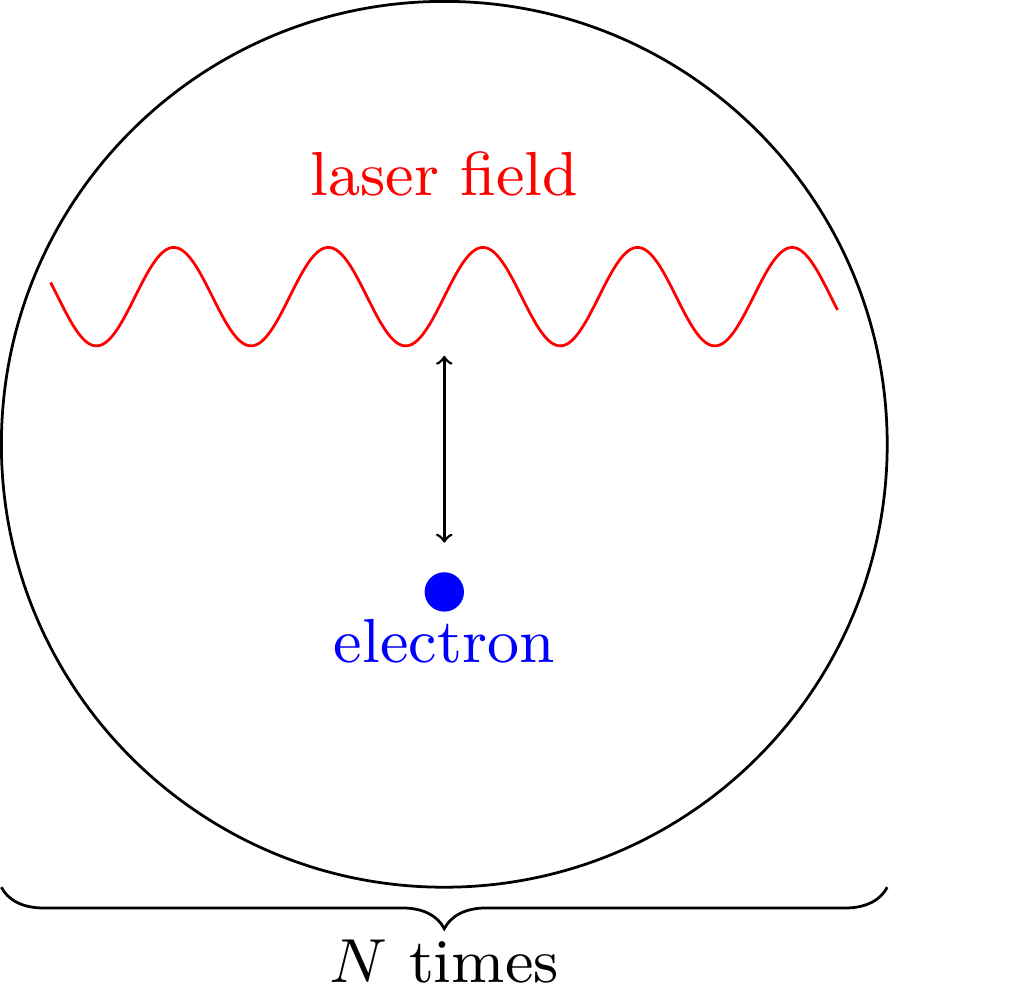}\includegraphics[width=0.2 \textwidth]{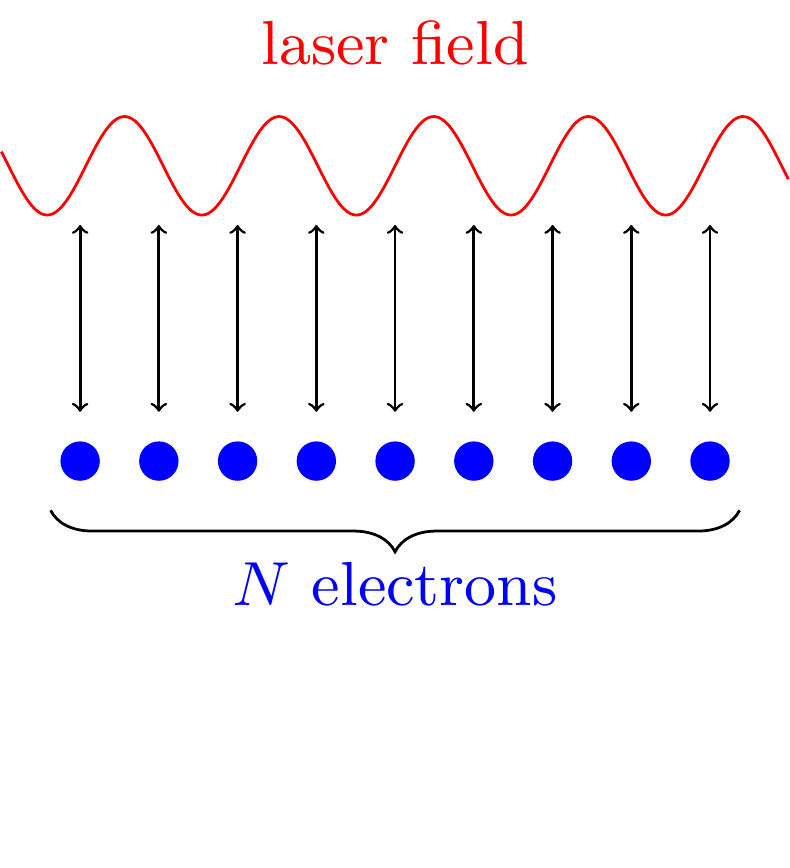}
\caption{Difference between the single-electron model for the low-gain regime (left) and the many-electron model for the high-gain FEL (right): In the low-gain regime we consider only a single electron interacting with the laser field and simply multiply the resulting change of the laser field with the number $N$ of electrons in the bunch. For the simultaneous interaction of all $N$ electrons with the laser field the motion of each electron influences the dynamics of the laser field which in turn acts back on the motion of all electrons. Hence, the electrons are indirectly coupled to each other due to their common interaction with the laser field.}  
\label{fig:rho}
\end{figure}

In Fig.~\ref{fig:rho} we have illustrated the difference between the single-electron (left) and the many-electron (right) approach. Our previous model is based on the interaction of one electron with the laser field and incorporates the effects of many electrons by the simple multiplication with the number $N$ of electrons in one bunch. 

In the high-gain regime on the right-hand side of Fig.~\ref{fig:rho}, we additionally take collective effects into account that emerge when all electrons simultaneously interact with the laser field: the motion of one particular electron leads to a change of the laser field which in turn influences the dynamics of the remaining electrons. In some sense, the electrons communicate with each other via the laser field. 

We emphasize that this collective effect must not be confused with the direct Coulomb interaction of the electrons due to space charge~\cite{sprangle1}, which is neglected here in analogy to the classical theory of a Compton FEL~\cite{boni_90}. 

\subsubsection{Entangled electron states}

The richer dynamics of the collective model becomes evident, when we regard the action of the Hamiltonian, Eq.~\eqref{eq:H_many}, on the electron motion. For a single electron, that is $N=1$, the momentum of the electron receives a kick by the recoil $q\equiv 2\hbar k$,
when a photon is absorbed from the laser field leading to the combined  action 
\ba\label{eq:scatt_bas}
\hat{a}_\L\e{\I2k\hat{z}}\ket{n,p}=\ket{n-1,p+q}
\ea 
of the momentum shift operator and the photon annihilation operator. Here we assumed that the electron initially is in a momentum eigenstate with momentum $p$. The laser field is described by a Fock state with photon number $n$, that is the eigenstate of the photon-number operator $\hat{n}\equiv \hat{a}_\L^\dagger \hat{a}_\L$ with the eigenvalue $n$. 

The behavior in Eq.~\eqref{eq:scatt_bas} has enabled us in Ref.~\cite{NJP2015} to expand the total state vector for the electron and the laser field in terms of the scattering basis~\cite{ciocci,*dattoli_hb4}, $\ket{\mu}\equiv \ket{n+\mu,p-\mu q}$. The single quantum number $\mu$ simultaneously corresponds to the number of scattered photons as well as to the change of the electron momentum as integer multiple of the recoil $q$.

\begin{figure}
\includegraphics[width=0.48 \textwidth]{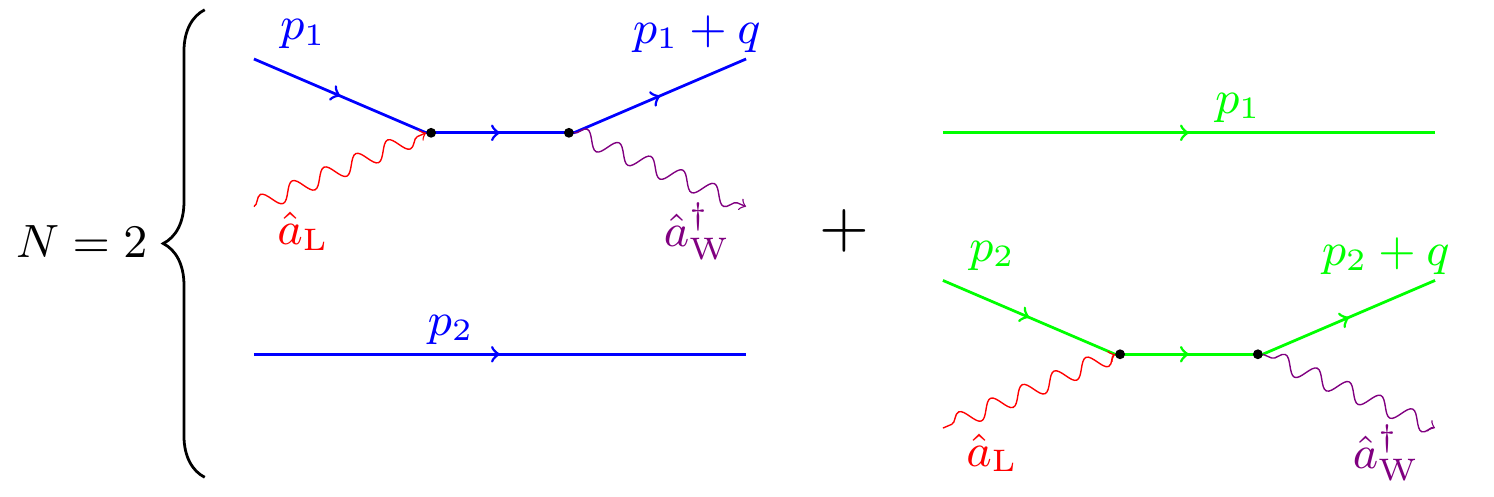}

\caption{Creation of entangled superposition states in the FEL for $N=2$ electrons: on the left-hand side, the first electron with momentum $p_1$ (above) absorbs a laser photon and emits a wiggler photon which leads to the increased momentum $p_1+q$ with $q\equiv 2\hbar k$ denoting the discrete recoil. In contrast, the second electron (below) does not scatter with the fields and thus maintains its initial momentum $p_2$. The second possible event is depicted on the right-hand side: here the first electron does not interact with the field and its momentum $p_1$ stays unchanged, while the second electron is scattered from a laser and a wiggler photon yielding the final momentum $p_2+q$. Superimposing these two possibilities leads us finally to the entangled state in Eq.~\eqref{eq:qhigh_N2}. For didactic reasons, we have used in this figure a representation, where besides the laser mode, described by $\hat{a}_\L$ and $\hat{a}_\L^\dagger$,  also the wiggler field is quantized with the photon annihilation and creation operators, $\hat{a}_\W$ and $\hat{a}_\W^\dagger$, respectively.}
\label{fig:two_paths}
\end{figure}

However, already for $N=2$ electrons the situation changes drastically: The sum in the Hamiltonian, Eq.~\eqref{eq:H_many}, over the different electrons leads to the entangled superposition state 
\begin{equation}\label{eq:qhigh_N2} 
\left(\e{\I 2k\hat{z}_1}+\e{\I 2k\hat{z}_2}\right)\ket{p_1,p_2}
=\ket{p_1+q,p_2}+\ket{p_1,p_2+q}
\end{equation}
for the electrons. Here we used  a product of two momentum eigenstates with eigenvalues $p_1$ and $p_2$, respectively, as initial state. 
The emergence of this superposition state becomes clear in Fig.~\ref{fig:two_paths}:
Either the first electron scatters from the laser and the wiggler field and receives a momentum kick, while the second electron is unaffected by the interaction, or the vice-versa process occurs, where the first electron maintains its initial momentum $p_1$ and the momentum of the second electron changes from $p_2$ to $p_2+q$ due to the interaction with the fields. By superimposing these two possible events we arrive at the state vector in Eq.~\eqref{eq:qhigh_N2}.  

From the inspection of Eq.~\eqref{eq:qhigh_N2} it becomes obvious that an expansion of the total state vector in terms of the scattering basis is not convenient since we do not know which electron has absorbed or emitted the photon. 

This statement is of course also true for the more general case of $N$ electrons, where we find the expression       
\begin{equation}\label{eq:qhigh_Narb}
 \begin{aligned}
    \sum\limits_{j=1}^N\e{\I 2k\hat{z}_j}\ket{p_1,p_2,...,p_N}
     =\sum\limits_{j=1}^N \ket{p_1,...,p_j+q,...,p_N}\,,
 \end{aligned}
\end{equation}
if each electron is initially in a momentum eigenstate. The form of this state vector is analogous to a Dicke state~\cite{dicke} in the field of superradiance and amplified spontaneous emission.

However, we emphasize that the dynamics of an FEL in general is richer than the one in the Dicke model, where the state of each atom is limited to two levels, which would correspond to two momenta $p_j$ and $p_j+q$ in Eq.~\eqref{eq:qhigh_Narb}. If the sum in Eq.~\eqref{eq:H_many} acts a second time on a product of $N=2$ momentum eigenstates  we obtain for example the expression  
\begin{equation}
 \begin{aligned}
     \left(\e{\I 2k\hat{z}_1}+\e{\I 2k\hat{z}_2}\right)^2\ket{p_1,p_2}
      = \ket{p_1+2q,p_2}+\ket{p_1,p_2+2q}\\ + 2\ket{p_1+q,p_2+q}\,,
 \end{aligned}
\end{equation}
which in contrast to the Dicke-like state from Eq.~\eqref{eq:qhigh_N2} includes three instead of only two momentum levels. Similarly,  we find for $N$ electrons the state 
\begin{equation}
 \begin{aligned}
   \left(\sum\limits_{j=1}^N\e{\I 2k \hat{z}_j}\right)^2\ket{p_1,p_2,..,p_N}=\sum\limits_{j=1}^N\ket{p_1,...,p_j+2q,...,p_N}\\
   +\sum\limits_{j\neq k}^N\ket{p_1,...,p_j+q,...,p_k+q,...,p_N}
 \end{aligned}
\end{equation}
which describes two-photon as well as single-photon processes. In the single-electron model, on the other hand, we find only $\ket{p+2q}$, that is a definite momentum change of $2q$.  

Through the successive action of the collective operator $\sum_j \exp{\left(\pm\I 2k \hat{z}_j\right)}$
we would arrive at even more involved expressions. Hence, we discard time-dependent state vectors in the Schr{\"o}dinger
picture. Instead, we study in the following the dynamics of operators in the Heisenberg picture.

\subsection{Collective operators}

Similarly to the collective variables~\cite{bnp,boni_coll} of the classical theory and of the quantum theory~\cite{boni06} for high-gain FELs,  we investigate the  dynamics of \emph{collective} operators.
However, there is a crucial difference in our treatment of the  Quantum FEL to the one in Ref.~\cite{boni06}. While the authors of Ref.~\cite{boni06} attempted to generalize the classical model of Refs.~\cite{bnp,boni_coll} by introducing a bunching operator  and a symmetrically ordered momentum bunching operator, we refine the ideas and concepts of our low-gain theory~\cite{NJP2015} of the Quantum FEL. For this purpose, we  introduce in the following collective momentum jump operators.   
 
\subsubsection{Definition} 
 
We identify the quantum regime of the FEL as the limit, where the infinite momentum ladder characterizing the electron dynamics reduces to only two resonant momentum levels. 

In order to describe the momentum ladder of each electron we introduce projection operators of the form     
\ba\label{eq:qhigh_proh} 
\hat{\Upsilon}_{\mu,\nu}\equiv \sum\limits_{j=1}^N\hat{\sigma}_{\mu,\nu}^{(j)}
\equiv\sum\limits_{j=1}^N \ket{p-\mu q}^{(j)}\bra{p-\nu q}\,.
\ea
Here we treat the momentum of the electron as a discrete variable, visualized by the integer multiples $\mu$ and $\nu$ of the recoil $q$, instead of a continuous one, since the Hamiltonian of Eq.~\eqref{eq:H_many} allows only for these discrete jumps. We therefore identify $\hat{\Upsilon}_{\mu,\nu}$ as a collective momentum jump operator, while 
\ba\label{eq:sigmamunu_def}
\sigma_{\mu,\nu}^{(j)}\equiv  \ket{p-\mu q}^{(j)}\bra{p-\nu q}   
\ea
describes the jump of electron $j$. 

Moreover, we note that the definition of  $\hat{\Upsilon}_{\mu,\nu}$ in Eq.~\eqref{eq:qhigh_proh} is only reasonable, if each electron has the same initial momentum $p_j\equiv p$ which corresponds to an initial state of the form $\ket{p,p,...,p}$.

\subsubsection{Operator algebra}

The commutation properties of the collective jump operators are described by the expression
\ba\label{eq:qhigh_comm}
\left[\hat{\Upsilon}_{\mu,\nu},\hat{\Upsilon}_{\rho,\eta}\right]
=\updelta_{\nu,\rho}\hat{\Upsilon}_{\mu,\eta}
-\updelta_{\eta,\mu}\hat{\Upsilon}_{\rho,\nu}
\ea
which follows from the corresponding single-electron relation
\ba\label{eq:qhigh_singlecomm} 
\left[\hat{\sigma}_{\mu,\nu}^{(j)},\hat{\sigma}_{\rho,\eta}^{(j)}\right]
=\updelta_{\nu,\rho}\hat{\sigma}_{\mu,\eta}^{(j)}
-\updelta_{\eta,\mu}\hat{\sigma}_{\rho,\nu}^{(j)}
\ea
and from the fact that operators corresponding to different electrons commute, that is $\left[\hat{\sigma}_{\mu,\nu}^{(j)},\hat{\sigma}_{\rho,\eta}^{(k)}\right]=0$ for $j\neq k$. We note that we have used Kronecker deltas
in Eqs.~\eqref{eq:qhigh_comm} and~\eqref{eq:qhigh_singlecomm}  instead of Dirac delta functions, since we assume discrete momentum ladders instead of a continuum and therefore $\braket{p-\mu q|p-\nu q}=\updelta_{\mu,\nu}$. 

The comparison of Eq.~\eqref{eq:qhigh_comm} to Eq.~\eqref{eq:qhigh_singlecomm} reveals that the single-electron jump operators and the collective operators commute  the same way. We emphasize that both sets of operators do not obey the same algebra: an important difference arises when we consider the product of two operators. For a single particle with $\hat{\sigma}_{\mu,\nu}\equiv\hat{\sigma}_{\mu,\nu}^{(j)} $ and $j=1$, we can write down the closed expression $\hat{\sigma}_{\mu,\nu}\hat{\sigma}_{\rho,\eta}=\updelta_{\nu,\rho}\hat{\sigma}_{\mu,\eta}$, which is impossible for the collective case, that is $\hat{\Upsilon}_{\mu,\nu}\hat{\Upsilon}_{\rho,\eta}\neq\updelta_{\nu,\rho}\hat{\Upsilon}_{\mu,\eta}$.

We exemplify this different behavior by the action of $\hat{\Upsilon}_{2,1}\hat{\Upsilon}_{1,0}$ on the state $\ket{p,p}$ for $N=2$ electrons. In the first step we obtain the expression
\ba\label{eq:agf} 
\hat{\Upsilon}_{2,1}\hat{\Upsilon}_{1,0}\ket{p,p}=
\hat{\Upsilon}_{2,1}\left(\ket{p-q,p}+\ket{p,p-q}\right)\,,
\ea
where the first operator $\hat{\Upsilon}_{1,0}$ creates an entangled state in analogy to the discussion for $\sum_j\exp{\left(\I 2k \hat{z}_j\right)}$ in the preceding section. In contrast to the single-electron case, where the action of $\hat{\sigma}_{1,0}$ on $\ket{p}$ yields a \textit{definite} momentum shift, that is the state $\ket{p-q}$, for two or more electrons the momentum of a particular electron is shifted only with a \textit{certain probability}. 

The second operator $\hat{\Upsilon}_{2,1}$ now acts on the entangled state in Eq.~\eqref{eq:agf} and we finally arrive at       
\ba 
\hat{\Upsilon}_{2,1}\hat{\Upsilon}_{1,0}\ket{p,p}=\ket{p-2q,p}+\ket{p,p-2q}\,.
\ea
This expression differs from the corresponding result $\hat{\sigma}_{2,1}\hat{\sigma}_{1,0}\ket{p}=\ket{p-2q}$ for a single electron and therefore constitutes a collective effect. 

\subsubsection{Rewriting the Hamiltonian}
    
We continue by transforming the original Hamiltonian from the Heisenberg picture such that the free part of the dynamics is included in the phases of the interaction term, analogously to the interaction picture. As a consequence, we identify  the important time scales of the FEL dynamics
directly in the Hamiltonian. For this purpose, the expansion of $\hat{H}$ in terms of the projection operators $\hat{\Upsilon}_{\mu,\nu}$ proves to be helpful.

This reformulation of the Hamiltonian and the transformation into the rotating frame are presented in detail in App.~\ref{app:Transformation_of_Hamiltonian_in_terms_of_jump_operator}. There, we derive the expression, Eq.~\eqref{eq:app_qhigh_HI},
\begin{equation}\label{eq:qhigh_HI}
\hat{H}'(\taub)=\varepsilon  \left(\hat{a}_\L \sum\limits_\mu \e{\I2\mu\taub} \hat{\Upsilon}_{\mu,\mu+1}
+\text{h.c.}\right) 
-\del\,\hat{n}
\end{equation}
for the Hamiltonian, where we have defined the
dimensionless coupling $\varepsilon \equiv g/\omega_\text{r}$  and the dimensionless time $\taub\equiv \omega_\text{r}t$. Moreover, we have recalled from Ref~\cite{NJP2015} the recoil frequency
\ba\label{eq:rec_freq} 
\omega_\text{r}\equiv \frac{1}{\hbar}\frac{q^2}{2m}
\ea
and  have introduced the relative deviation 
\ba\label{eq:delta_def} 
\del\equiv \frac{p-q/2}{q/2}
\ea
of the initial momentum $p$ of the electrons from $p=q/2$.  The deviation $\del$ will play the role of a detuning since we later identify $q/2$ with the resonant momentum in the quantum regime. 

Because we consider time-dependent operators, we have to solve the Heisenberg equation of motion, Eq.~\eqref{eq:app_eqofmot},
\ba\label{eq:hber_eqmot_ht}
\I\frac{\D}{\D\taub}\hat{\mathcal{O}}'(\taub)=\left[\hat{\mathcal{O}}'(\taub),\hat{H}'(\taub)\right]\,
\ea
subject to the Hamiltonian $\hat{H}'$ in Eq.~\eqref{eq:qhigh_HI}, in order to obtain the time evolution of an operator $\hat{\mathcal{O}}'$.

The comparison of $\hat{H}'$ in Eq.~\eqref{eq:qhigh_HI} to $\hat{H}$ in Eq.~\eqref{eq:H_many} reveals that we do not sum over the different electrons any more, but over different momenta
which are integer multiples of  the recoil $q$. The collective effect of all electrons on the laser field is now fully contained in the jump operators $\hat{\Upsilon}_{\mu, \nu}$. 

\subsection{Dicke Hamiltonian}

In Ref.~\cite{NJP2015} 
we have found that for a large recoil multiphoton transitions are suppressed and only the two resonant momentum levels $p=q/2$ and $p=-q/ 2$ are of importance similarly to atomic Bragg diffraction~\cite{giese}. This reduction to the interaction of a two-level system with a quantized field mode has led us to the analogy of the Quantum FEL to the Jaynes-Cummings model~\cite{jc}.

The natural generalization for the many-particle case is the Dicke model~\cite{dicke}, where many two-level atoms  simultaneously interact with the
field mode. In the following we establish this analogy for the many-electron model. 

\subsubsection{Relevant time scales}

We first separate the slowly-varying dynamics from  the rapid oscillations.
For that we make use of the canonical variant~\citep{higher} of the method of averaging, in which we directly work with the Hamiltonian instead of the equations of motion. To employ this technique, we decompose $\hat{H}'(t)$, Eq.~\eqref{eq:qhigh_HI}, into a Fourier series
\ba\label{eq:qhigh_Hfour} 
\hat{H}'(\taub)\equiv \varepsilon \sum\limits_\mu \hat{\mathcal{H}}_\mu \e{\I2\mu \taub}
\ea
with
\ba\label{eq:qhigh_Fcomp} 
\begin{cases}
\hat{\mathcal{H}}_0\equiv \hat{a}_\L \hat{\Upsilon}_{0,1} +
\hat{a}_\L^\dagger \hat{\Upsilon}_{1,0}-\frac{\del}{\varepsilon} \hat{n}\\
\hat{\mathcal{H}}_\mu\equiv \hat{a}_\L \hat{\Upsilon}_{\mu,\mu+1} +
\hat{a}_\L^\dagger \hat{\Upsilon}_{-\mu+1,-\mu}
\end{cases}
\ea
denoting the Fourier components.

The inspection of the Fourier series in Eq.~\eqref{eq:qhigh_Hfour} reveals that there are contributions which are oscillating with multiples of the recoil frequency, that is $2\mu \taub=2\mu\omega_\text{r}t$. In addition, the component $\hat{\mathcal{H}}_0$ is independent of time. If the recoil is large, the oscillations with the recoil frequency are rapidly-varying terms and we neglect them in a rotating wave-like approximation~\cite{schleich}.   
Hence, the dynamics is determined by the time-independent part  
\ba\label{eq:appr_dicke}
\hat{H}_\text{eff}=\varepsilon \hat{\mathcal{H}}_0=\varepsilon\left(\hat{a}_\L \hat{\Upsilon}_{0,1} +
\hat{a}_\L^\dagger \hat{\Upsilon}_{1,0}\right)-\del\,\hat{n}
\ea
which we identify with the effective Hamiltonian in  lowest order of the method of averaging~\cite{higher}, that is $\hat{H}_\text{eff}\equiv\varepsilon \hat{H}_\text{eff}^{(1)}$.   
If we interpret $\hat{\Upsilon}_{1,0}$ and $\hat{\Upsilon}_{0,1}$ as collective spin-flip operators, we observe that the expression for the effective Hamitonian in Eq.~\eqref{eq:appr_dicke} is indeed equivalent to the Dicke Hamiltonian with an additional detuning $\del$~\footnote{In Ref.~\cite{gaiba} the analogy of the quantum regime to the Dicke Hamiltonian was also pointed out, however, within a different approach. There, the formalism of Ref.~\cite{preparata88} in second quantization was employed leading in the quantum regime to a Hamiltonian which is trilinear in bosonic operators. By means of the Schwinger representation~\cite{schwinger} of angular momentum, we can establish the connection of our model to the one in Ref.~\cite{gaiba}. In addition, we present in our article a rigorous proof for the two-level approximation with the help of the method averaging which goes beyond the perturbative approach in Ref.~\cite{gaiba}. Moreover, we have included a nonzero deviation $\del$ from resonance.}.

The dynamics of each electron in the quantum regime is thus described by the resonant scattering from the excited state, in the vicinity of the resonant momentum $p=q/2$, to the ground state $p\cong -q/2$.
All two-level systems collectively interact with the radiation field.

\subsubsection{Conditions for quantum regime}

We have to make sure that the approximation leading to Eq.~\eqref{eq:appr_dicke} is allowed. For that, the corresponding asymptotic expansion has to converge when we consider higher orders of the method of averaging. Therefore, we require that $\varepsilon|\hat{\mathcal{H}}_\mu|$ is small~\cite{higher}, that is $\varepsilon |\hat{\mathcal{H}}_\mu|\ll 1$. Estimating $ |\hat{\mathcal{H}}_\mu|\sim\sqrt{N}$ we obtain the fundamental condition $\alpha_N \equiv \varepsilon \sqrt{N}\ll 1$ for the Quantum FEL, where we have recalled the quantum parameter $\alpha_N$~\cite{NJP2015}.

This parameter describes the ratio 
\ba\label{eq:alpha} 
\alpha_N= \frac{g\sqrt{N}}{\omega_\text{r}}\ll 1
\ea
of the two important frequency scales of the FEL dynamics, that is the coupling strength $g\sqrt{N}$ and the recoil frequency $\omega_\text{r}$, Eq.~\eqref{eq:rec_freq}. Thus, we are in the quantum regime, if the recoil exceeds the coupling. 

Moreover, we require a small deviation $\del$ from resonance, that is $\del \ll 1 $, as well as a small initial photon number. Else, the magnitude of the second term of $\hat{H}_\text{eff}$ would be larger than the first term which scales with $\alpha_N$ and we cannot perform  an asymptotic expansion in powers of the quantum parameter.

In reality, the ensemble of electrons is not described by the same initial momentum, but each electron has a different one. Hence, the electrons move on momentum ladders which are shifted with respect to each other due to these different starting points. 
The different initial momenta $p_1,p_2,...,p_N$ are distributed according to a statistics which is characterized by a certain characteristic width $\Delta p$.
A small deviation $\del$ from resonance thus translates to a constraint for the width $\Delta p$ of the momentum distribution. From $\del < 1$ we deduce the requirement
\ba\label{eq:deltap_fund} 
\Delta p<q
\ea 
for realistic electron beams in analogy to Refs.~\cite{pio,NJP2015}.


%% file: Exponential_gain.tex
\section{Exponential gain}
\label{sec:Exponential_gain}

A key feature of a classical FEL in the  high-gain regime is the exponential growth of the laser intensity for short times~\cite{bnp,schmueser}. By solving the dynamics of an FEL in the deep quantum regime, dictated by the Dicke Hamiltonian within a parametric approximation~\cite{siegman,kumar}, we obtain this high-gain behavior also for a Quantum FEL. 


\subsection{Deep quantum regime}

In the following, we derive expressions for the mean photon number, the gain length, the gain function, and for the variance of the photon number of a high-gain Quantum FEL in the exponential-gain regime. For the time being we restrict ourselves to the two-level approximation, that is the Dicke Hamiltonian from Eq.~\eqref{eq:appr_dicke}.

\subsubsection{Parametric approximation}

In analogy to the Pauli spin matrix  $\hat{\sigma}_z$ we define the operator
\ba 
\hat{\Upsilon}_z\equiv \hat{\Upsilon}_{0,0}-\hat{\Upsilon}_{1,1}
\ea
as difference of $\hat{\Upsilon}_{0,0}$ corresponding to the excited state $p\cong q/2$ and $\hat{\Upsilon}_{1,1}$
corresponding to the ground state $p\cong-q/2$. If all electrons are in the excited state, the expectation value of $\hat{\Upsilon}_z$ is maximized and simply equals the number $N$ of electrons, that is $\braket{\hat{\Upsilon}_z}=N$. In contrast, if all electrons are in the ground state the mean value is minimized to $\braket{\hat{\Upsilon}_z}=-N$. Thus, we can interpret 
$\hat{\Upsilon}_z$ as an effective inversion operator.

For short interaction times, only a few electrons have changed from the excited state to the ground state. Hence, we assume that $\hat{\Upsilon}_z$ has not changed very much and we can treat it as a constant. This procedure is known as `parametric approximation'~\cite{kumar} since we obtain the equations of motion for a parametric amplifier~\cite{siegman}.

To perform this approximation, we first scale the involved jump operators in the following way 
\ba\label{eq:qhigh_tilde} 
\begin{cases}
\hat{\mathcal{Y}}_{1,0}&\equiv \frac{1}{\sqrt{N}}\hat{\Upsilon}_{1,0}\\
\hat{\mathcal{Y}}_{z}&\equiv \frac{1}{N}\hat{\Upsilon}_{z}\\
\end{cases}
\ea
where $\hat{\mathcal{Y}}_z$ is bounded by $-1\leq \hat{\mathcal{Y}}_z \leq 1 $.

The Heisenberg equations of motion from Eq.~\eqref{eq:hber_eqmot_ht} of the rescaled operators read
\begin{subequations}
   \begin{align}
\label{eq:qhigh_nlina}
\I \frac{\D}{\D\taub}\hat{\mathcal{Y}}_{1,0}&= -\alpha_N \hat{a}_\L \hat{\mathcal{Y}}_z  \\\label{eq:qhigh_nlin}
\I \frac{\D}{\D\taub}\hat{\mathcal{Y}}_{z} &=\frac{\alpha_N}{N/2} 
\left(\hat{a}_\L \hat{\mathcal{Y}}_{0,1}
-\hat{a}_\L^\dagger \hat{\mathcal{Y}}_{1,0} \right)\\
\label{eq:qhigh_nlinc}
\I \frac{\D}{\D\taub}\hat{a}_\L &=- \del \hat{a}_\L+ \alpha_N \hat{\mathcal{Y}}_{1,0}\,,
   \end{align}
\end{subequations}
where we have recalled the quantum parameter $\alpha_N$ from Eq.~\eqref{eq:alpha} and have  used the commutation relation from Eq.~\eqref{eq:qhigh_comm} for the jump operators as well as the one for the laser-field  operators. 

Moreover, we assume that the initial state of the combined system of laser field and electrons in the rotating frame is given by the product    
\begin{equation}\label{eq:qhigh_in} 
\ket{\Psi'(\taub)}=\e{-\I\del n_0 \taub}\e{-\I N(\del+1/2)^2\taub}\ket{n_0}\otimes \ket{p,p,...,p}
\end{equation}
consisting of a Fock state with $n_0$ laser photons and of momentum eigenstates for the electrons. Here each electron has the same initial momentum $p$~\cite{becker83,fussnote1} 
that represents the excited state in vicinity of the resonance $p=q/2$. In addition, the photon number should be much smaller than the number of electrons, that is $n_0\ll N$. Else the following linearization procedure would break down since  $\hat{a}_\L$ is not small when compared to $\hat{\mathcal{Y}}_z$. We note that the additional phase factors in Eq.~\eqref{eq:qhigh_in} have emerged since we have performed a transformation into a time-dependent picture according to Eq.~\eqref{eq:app_qhigh_trafo2}. 
 
We now apply the parametric approximation by assuming that $\displaystyle\hat{\mathcal{Y}}_z$ is constant and replacing it by its expectation value at $\tau=0$, which is given by $\braket{\hat{\mathcal{Y}}}_z=1$. Hence, we arrive at the linear set of differential equations
\ba\label{eq:qhigh_lin_deep} 
\I \frac{\D}{\D\taub}
\begin{pmatrix}
\hat{\mathcal{Y}}_{1,0}\\
\hat{a}_\L
\end{pmatrix}=
\begin{pmatrix}
0 & -\alpha_N\\
 \alpha_N &-\del 
\end{pmatrix}
\begin{pmatrix}
\hat{\mathcal{Y}}_{1,0}\\
\hat{a}_\L
\end{pmatrix}
\ea
for the dynamics of electrons and laser field in a high-gain Quantum FEL.

If only a comparatively small number of photons is emitted,  the right-hand side of the original equation of motion, Eq.~\eqref{eq:qhigh_nlin}, for $\hat{\mathcal{Y}}_z$ is suppressed with $1/N$.
In this case, we are allowed to use the linearized equation of motion in Eq.~\eqref{eq:qhigh_lin_deep}.

However, for longer times the photon number $n$ grows and the parametric approximation breaks down~\cite{kumar}:
At some point we have $n\sim N$ and the right-hand side of Eq.~\eqref{eq:qhigh_nlin} is of the order of $\alpha_N$. Hence, the rate for the change of $\hat{\mathcal{Y}}_z$ scales the same as for $\hat{a}_\L$ and $\hat{\mathcal{Y}}_{1,0}$, Eqs.~\eqref{eq:qhigh_nlina} and~\eqref{eq:qhigh_nlinc}, and thus we cannot approximate $\hat{\mathcal{Y}}_z$ as a constant any longer. 
We postpone the investigation of this long-time behavior to a future article and restrict ourselves here to the linearized dynamics.

\subsubsection{Mean photon number \& gain length}

The linearized differential equation, Eq.~\eqref{eq:qhigh_lin_deep}, can be straightforwardly solved with the ansatz $\sim \e{-\I\lambda \tau}$. This procedure leads to two solutions 
\ba\label{eq:qhigh_lambda_deep} 
\lambda_\pm =-\frac{\del}{2}\pm \I \alpha_N\sqrt{1-\frac{\varkappa^2}{4}}
\ea
of the resulting quadratic equation. Here $\varkappa\equiv \del/\alpha_N $ denotes the deviation from resonance normalized to the quantum parameter $\alpha_N$. Since we assume $p$ in the vicinity of the resonant momentum $p=q/2$, quantified by the condition  $\del\equiv\varkappa \alpha_N\ll 1$, we require that $\varkappa$ is maximally of the order of unity, that is $\varkappa \sim \mathcal{O}(1)$. 
\begin{figure}
\centering
\includegraphics[width=0.45 \textwidth]{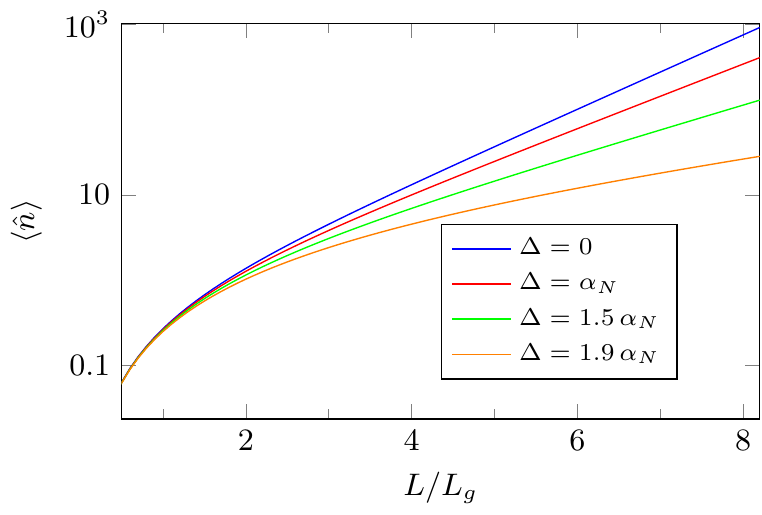}



\caption{Exponential gain of a high-gain Quantum FEL for short times: we have drawn the mean photon number $\braket{\hat{n}}=n_\text{sp}$ emerging from spontaneous emission, Eq.~\eqref{eq:qhigh_ndeep}, as a function of the wiggler length $L$ in multiples of the gain length $L_g$, Eq.~\eqref{eq:qhigh_Lgain}. We, moreover, study the behavior of $\braket{\hat{n}}$ for four different values of the deviation $\del$, Eq.~\eqref{eq:delta_def}, from resonance. Besides the start-up from vacuum, that is from $\braket{\hat{n}(0)}=0$, we observe that the growth of the mean photon number is decreased for increasing values of $\del$.}
\label{fig:qhigh_exp}
\end{figure}

The inspection of Eq.~\eqref{eq:qhigh_lambda_deep} reveals that  $\lambda$ possesses a nonzero imaginary part 
for $-2<\varkappa <2$. Hence, we expect that the field grows exponentially in time with the increment $\text{Im}\lambda_+$. For resonance, this growth is characterized by $\alpha_N \taub \equiv L/(2L_g)$, where $L\equiv ct$ denotes the wiggler length while $L_g$ describes the typical length scale of the gain.

After solving Eq.~\eqref{eq:qhigh_lin_deep} for $\hat{a}_\L$ and $\hat{\mathcal{Y}}_{1,0}$ we can calculate expectation values of the involved operators with respect to the initial state in Eq.~(\ref{eq:qhigh_in}). For example, we obtain the expression 
\ba\label{eq:mean_exp} 
\braket{\hat{n}(L)}=(n_\text{sp}(L)+1)\braket{\hat{n}(0)}+n_\text{sp}(L)
\ea
for the expectation value of the photon-number operator $\hat{n}\equiv \hat{a}_\L^\dagger \hat{a}_\L$ as a function of  the wiggler length  $L$~\footnote{To calculate the expectation value in Eq.~(\ref{eq:mean_exp}) with respect to the initial state from Eq.~(\ref{eq:qhigh_in}), we have used the identity $\braket{\hat{\mathcal{Y}}_{0,1}(0)\hat{\mathcal{Y}}_{1,0}(0)}=1$ which is valid, if all electrons are in the excited state. All other occurring expectation values of jump operators $\hat{\mathcal{Y}}_{\mu,\nu}(0)$ are zero for this particular initial state.}.

If the field starts from vacuum, that is $\braket{\hat{n}(0)}=0$, only the term 
\ba\label{eq:qhigh_ndeep} 
n_\text{sp}(L)=\frac{1}{1-\varkappa^2/4}
\sinh^2{\left[\frac{L}{2L_g}\sqrt{1-\frac{\varkappa^2}{4}}\right]}
\ea
corresponding to \textit{spontaneous} emission is present.
For a seeded FEL, however, the first term in Eq.~\eqref{eq:mean_exp} dominates.
Since this contribution is proportional to the initial number of photons $\braket{\hat{n}(0)}$, it describes \textit{stimulated} emission. 

In  Fig.~\ref{fig:qhigh_exp} we have drawn $\braket{\hat{n}}$ against $L$ for different values
of $\del$. For this purpose, we have restricted us to the start-up from  vacuum. Indeed, we observe an exponential growth of the photon number. The typical length scale of this growth, at least for resonance $\varkappa=0$, is given by the gain length~\footnote{If we assume $L \ll L_g$, we observe that the exponential behavior in Eq.~(\ref{eq:qhigh_ndeep}) reduces to the quadratic dependency $\braket{\hat{n}}\sim (gT)^2N$ 
corresponding to the low-gain, small-signal limit~\cite{NJP2015} of a Quantum FEL emerging from ordinary perturbation theory.} 
\ba\label{eq:qhigh_Lgain} 
L_g\equiv \frac{c}{2g\sqrt{N}}
\ea
which is consistent with the expression for $L_g'$ in Ref.~\cite{boni06}.

We recognize that the gain length $L_g$ in the quantum regime differs from the corresponding classical quantity~\cite{schmueser}, which reads
\ba\label{eq:Lgain_cl}
L_g^\text{(cl)}\equiv\frac{1}{\sqrt{3}} \frac{c}{(g^2N\omega_\text{r}/2)^{1/3}},
\ea
in the Bambini--Renieri frame~\cite{peter} and which emerges by solving a cubic characteristic equation~\citep{bnp}. The comparison of $L_g$ from Eq.~\eqref{eq:qhigh_Lgain}  to $L_g^{(\text{cl})}
$ from Eq.~\eqref{eq:Lgain_cl} yields the relation
\ba 
\frac{L_g}{L_g^{(\text{cl})}}=\frac{\sqrt{3}}{2 ^{4/3}}\frac{1}{\alpha_N^{1/3}}\,.
\ea 
Due to $\alpha_N\ll 1$ the gain length in the quantum regime is longer than predicted by a classical theory which can be interpreted
 as a quantum effect.

\subsubsection{Gain function}

Coming back to Fig.~\ref{fig:qhigh_exp}, we do not only observe that the mean photon number grows exponentially, but also that this growth decelerates when we move away from resonance, that is for increasing values of $\del$.

\begin{figure}
\centering
\includegraphics[width=0.45 \textwidth]{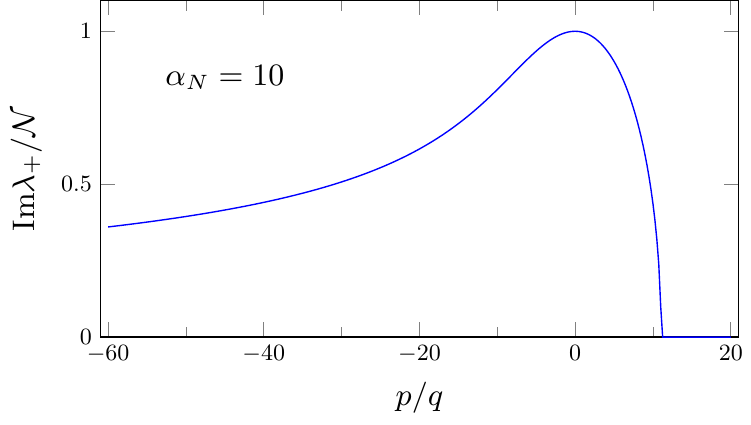}
\includegraphics[width=0.45 \textwidth]{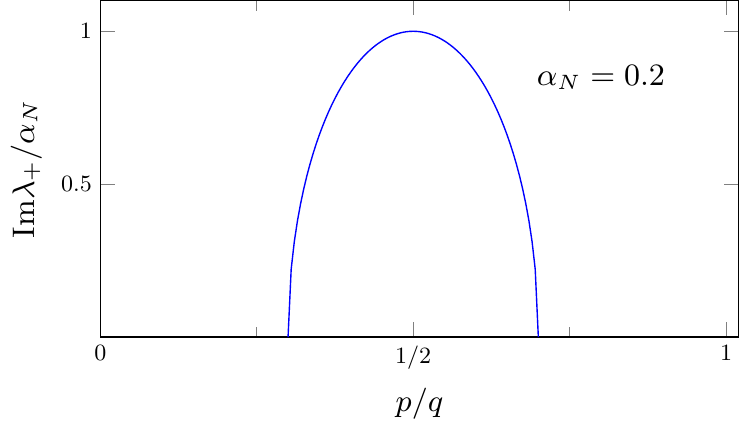}
\caption{Gain functions in the classical and in the quantum regime: we have drawn the increments $\text{Im}\lambda_+$ for the intensity growth in a classical FEL (top) and in a Quantum FEL (bottom), respectively, both as functions of the momentum $p$ divided by the recoil $q$. For the former case we numerically solve the cubic characteristic equation from Ref.~\cite{bnp} for $\alpha_N=10$, while we use our analytic expression from Eq.~\eqref{eq:qhigh_imlambda_deep} for the latter case with $\alpha_N=0.2$. The classical curve reaches its maximum at $p=0$ and is a smooth and broad function that covers many multiples of $q$. In contrast, its quantum counterpart is sharply peaked at the quantum resonance, $p=q/2$, and is very narrow, covering a range of momenta which is smaller than $q$. Hence, the gain bandwidth in a Quantum FEL is much smaller than in a classical FEL.}
\label{fig:high_cl_q}
\end{figure}

We quantify this effect by the positive imaginary part~\cite{qfelii}  
\ba\label{eq:qhigh_imlambda_deep}
\text{Im}\lambda_+ =\alpha_{N}\sqrt{1-\frac{\varkappa^2}{4}}
\ea
of $\lambda$ which is half the increment of $\braket{\hat{n}}$. We recognize from Eq.~\eqref{eq:qhigh_imlambda_deep} that for increasing values of $\varkappa$ the growth of 
$\braket{\hat{n}}$ decreases while it is maximized for resonance, $\varkappa=0$. Hence, we identify $\text{Im}\lambda_+$ as the gain function of a high-gain Quantum FEL. This definition is analogous to the classical regime, with the difference that there $\text{Im}\lambda_+$ emerges by solving a cubic characteristic equation~\cite{bnp} instead of a quadratic one~\cite{qfelii,fares2018} in the quantum regime.

In Fig.~\ref{fig:high_cl_q}  we compare the gain function of a classical high-gain FEL (top)  with the one, Eq.~\eqref{eq:qhigh_imlambda_deep}, of a Quantum FEL (bottom), both drawn against the initial momentum $p$ of the electrons. While the classical gain is maximized at $p=0$, the maximum gain in the quantum domain is located at $p=q/2$. Moreover, we observe that the gain function in the  classical case is a smooth curve which covers a wide range of momenta over many multiples of the recoil $q$. In contrast, the Quantum FEL is characterized by a sharp resonance with a width in momentum space that is smaller than $q$.

The gain curve is different from zero for $-2< \varkappa < 2$, which corresponds to a width of $2\alpha_N q$ in momentum space. We identify this width of the gain function as the gain bandwidth of a high-gain Quantum FEL~\cite{pio}, since only electrons with momenta that are within this region resonantly interact  with the fields, an effect which is known as velocity selectivity~\cite{giese,giltner,szigeti}. For an electron beam with the initial momentum spread $\Delta p$ we thus deduce the condition 
\ba\label{eq:gain_bw} 
\Delta p < 2\alpha_N\,q
\ea
for efficiently amplifying the laser field. We note that this requirement is stricter than the fundamental one $\Delta p < q$ from Eq.~\eqref{eq:deltap_fund} due to $\alpha_N\ll 1$.

We emphasize that an analogous behavior of the gain function is discussed in Refs.~\cite{boni06,bonifacio-basis}. However, we here present a simple interpretation through the analogy to the Dicke-Hamiltonian. Moreover, we write down an explicit analytic expression  in Eq.~\eqref{eq:qhigh_imlambda_deep} for the gain function.

\subsubsection{Variance of photon number}

Solving the Heisenberg equation of motion gives us the time dependency of the field in terms of the operator $\hat{a}_\L$. Hence, this solution enables us not only to  calculate the mean photon number  $\braket{\hat{n}}$ but also its higher moments.  

As an example, we consider in the following the variance  
\ba\label{eq:qhigh_sigma} 
\Delta n^2(L)\equiv \braket{\hat{n}^2(L)}-\braket{\hat{n}(L)}^2
\ea
of the photon number. 

We straightforwardly derive the expression    
\begin{equation}\label{eq:qhigh_chaot} 
\Delta n^2(L)=\left(n_\text{sp}(L)+1\right)^2\Delta n^2(0)+n_\text{sp}(L)\braket{\hat{n}(L)+1}
\end{equation}
in terms of the initial variance $\Delta n^2(0)$, where we have recalled  $\braket{\hat{n}}$ and $n_\text{sp}$ from Eqs.~\eqref{eq:mean_exp} and~\eqref{eq:qhigh_ndeep}, respectively. Since $n_\text{sp}$ grows exponentially and all terms in Eq.~\eqref{eq:qhigh_chaot} are larger than or equal zero, the variance $\Delta n^2$ becomes larger than the mean value $\braket{\hat{n}}$, that is  $\Delta n^2>\braket{\hat{n}}$. 
Hence, we obtain a super-Poissonian  photon statistics for a high-gain Quantum FEL in the exponential-gain regime which even moves farther away from a Poissonian behavior with $\Delta n^2=\braket{\hat{n}}$ for increasing values of $L$.  

\begin{figure}
\centering
\includegraphics[width=0.46 \textwidth]{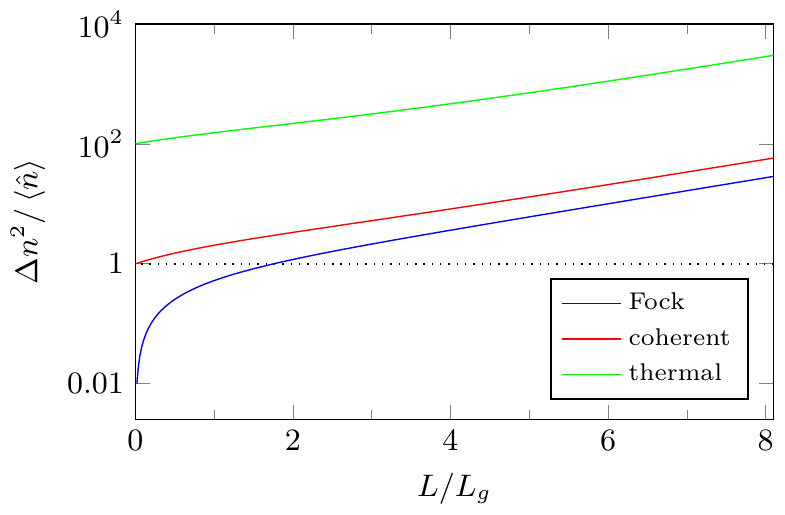}






\caption{Variance $\Delta n^2$, Eq.~\eqref{eq:qhigh_chaot}, of the photon number normalized to the expectation value $\braket{\hat{n}}$, Eq.~\eqref{eq:mean_exp}, as a function of the length $L$ of the wiggler in units of the gain length $L_g$, Eq.~\eqref{eq:qhigh_Lgain}, for exact resonance $p=q/2$. We compare the situation, where the field is initially in (i) a Fock state (blue line), (ii) a coherent state (red line), and (iii) a thermal state (green line), with each of them being described by the same mean photon number, that is $\braket{\hat{n}(0)}=100$. In all cases we observe a super-Poissonian behavior of the photon statistics which moves away from a Poisson statistics with $\Delta n^2=\braket{\hat{n}}$ (dotted line) for increasing values of $L$.
We find the greatest deviation from a Poissonian for an initially thermal field, while the case of an initial Fock state shows the smallest deviation.}
\label{fig:qhigh_exp_var}
\end{figure}

If the field starts from vacuum, where $\braket{\hat{n}(0)}=0$, we find the expression
\ba\label{eq:thermal_asympt} 
\Delta n^2(L)=\braket{\hat{n}(L)}\braket{\hat{n}(L)+1}
\ea
that is the field obeys thermal statistics~\cite{siegman,boniprep70}. In addition, the relation in Eq.~\eqref{eq:thermal_asympt}  holds true, if the field starts in a thermal state, that is a thermal state maintains its nature during interaction.     

On the other hand, we derive for an initial Fock state the asymptotic behavior 
\ba\label{eq:fock_var} 
\frac{\Delta n^2(L)}{\braket{\hat{n}(L)}}\cong\frac{\braket{\hat{n}(L)}+1}{\braket{\hat{n}(0)}+1}\,,
\ea
for $L\rightarrow \infty$. Since $\braket{\hat{n}}$ increases with $L$ we easily deduce from Eq.~\eqref{eq:fock_var}  a super-Poissonian behavior of the laser field. However, for relatively large values of the initial photon number $\braket{\hat{n}(0)}$ the field  clearly deviates  from a thermal state like in Eq.~\eqref{eq:thermal_asympt}. 

Indeed, we observe in Fig.~\ref{fig:qhigh_exp_var} that a field which is initially described by a thermal state clearly deviates more from a Poissonian statistics with $\Delta n^2=\braket{\hat{n}}$ than the corresponding  case of a Fock state. If the field is initially in a coherent state, we find that the normalized variance $\Delta n^2/\braket{\hat{n}}$ lies in between the two extremes of a thermal and a Fock state, respectively.

\subsection{Higher-order corrections}

So far, we have focused on the deep quantum regime defined by the the Dicke Hamiltonian, Eq.~\eqref{eq:appr_dicke}. However, in order to prove that this two-level approximation is valid, we consider corrections to this limit which have to be suppressed for $\alpha_N\ll 1$. 
In the following, we perform this proof by calculating higher orders of the method of canonical averaging~\cite{higher} and by linearizing the resulting equations of motion in the exponential-gain regime. Moreover, we connect to the results  of Ref.~\cite{boni06}. 

\subsubsection{Canonical averaging}

We  present here only the main ideas and results of our approach. However, the interested reader may find the details of these calculations in App.~\ref{app:Canonical_averaging}. There, we show that the dynamics of an operator $\hat{\mathcal{O}}'$ is dictated by the equation of motion    
\ba
\I \frac{\text{d}}{\text{d}\tau}\hat{\mathcal{O}}'=\varepsilon\left[\hat{\mathcal{O}}',\hat{H}^{(1)}_\text{eff}\right]+\varepsilon^2\left[\hat{\mathcal{O}}',\hat{H}^{(2)}_\text{eff}\right]
+\varepsilon^3\left[\hat{\mathcal{O}}',\hat{H}^{(3)}_\text{eff}\right]\,,
\ea
up to third order in $\varepsilon$. We identify the lowest-order contribution $\hat{H}_\text{eff}^{(1)}$  with the Dicke Hamiltonian, Eq.~\eqref{eq:appr_dicke}, of the deep quantum regime, while the higher-order terms $\hat{H}_\text{eff}^{(2)}$ and $\hat{H}_\text{eff}^{(3)}$ are given by Eqs.~\eqref{eq:app_many_H2} and~\eqref{eq:app_many_H3}, respectively. 

In short, this asymptotic expansion emerges by separating the slowly-varying dynamics from the rapid oscillations and expanding the occurring terms in powers of $\varepsilon$. The effective Hamiltonian is then determined by considering in each order only contributions that are independent of time $\taub$, in order to avoid secular terms~\cite{higher}. We note that the rapid oscillations, neglected in lowest order, do have an averaged influence in the higher-order contributions of $\hat{H}_\text{eff}$ which emerge from cross terms, where the time-dependent phases cancel.       

\subsubsection{Linearization}

Similarly to the parametric approximation in the deep quantum regime, we linearize the equations of motion with the Hamiltonian from Eqs.~\eqref{eq:appr_dicke},~\eqref{eq:app_many_H2} and~\eqref{eq:app_many_H3}, by setting $\hat{\Upsilon}_{0,0}\cong N$ and by considering only contributions which are linear in $\hat{a}_\L\cong\updelta\hat{a}_\L$ and in $\hat{\Upsilon}_{\mu,\nu}\cong\updelta\hat{\Upsilon}_{\mu,\nu}$, except for $\mu=\nu=0$. This procedure, described in App.~\ref{app:Canonical_averaging}, yields the linearized set of equations
\ba\label{eq:qhigh_lin_corr}
\I\frac{\text{d}}{\text{d}\tau}
 \begin{pmatrix}
  \updelta{\hat{\mathcal{Y}}}_{1,0} \\ \updelta\hat{a}_\text{L}
 \end{pmatrix}=M \begin{pmatrix}
  \updelta{\hat{\mathcal{Y}}}_{1,0} \\ \updelta\hat{a}_\text{L}
 \end{pmatrix}\,,
\ea
with the matrix, Eq.~\eqref{eq:app_lin_corr},
\ba 
M\equiv\begin{pmatrix}
 0 & -\alpha_N\left(1-\frac{\alpha_N^2}{8}\right)\\
\alpha_N\left(1-\frac{\alpha_N^2}{8}\right) & -\alpha_N \left(\varkappa +\frac{\alpha_N}{2} -\frac{\varkappa\alpha_N^2}{4}\right)
\end{pmatrix}
\ea
coupling the dynamics of $\updelta\hat{a}_\L$ to the one of
 $\updelta\hat{\mathcal{Y}}_{1,0}\equiv\updelta\hat{\Upsilon}_{1,0}/\sqrt{N}$.

By assuming again a solution of the form $\sim\e{-\I\lambda \taub}$ we arrive at a quadratic equation for $\lambda$ which is straightforwardly solved. The positive imaginary part of this solution reads
\ba\label{eq:qhigh_lambda_corr} 
\text{Im}\lambda_+ \cong\alpha_N \sqrt{1-\frac{\varkappa^2}{4}}&\left[1 - \frac{\varkappa/2}{1-\frac{\varkappa^2}{4}}\frac{\alpha_N}{4}\right.\\
&\ \ \ \ \left.-\frac{5-3\varkappa^2+\varkappa^4/2}{\left(1-\frac{\varkappa^2}{4}\right)^2}\frac{\alpha_N^2}{32}\right]\,,
\ea
where we have kept only terms up to third order in $\alpha_N$. We note that going to second order of the expansion, Eq.~\eqref{eq:qhigh_lambda_corr}, would not be sufficient to observe corrections for the resonant case $\kappa=0$.
The first nonzero term emerges in third order and scales with $5\alpha_N^2/32$ compared to  first order.

\subsubsection{Discussion of results}

Indeed, we recover in Eq.~\eqref{eq:qhigh_lambda_corr} the result from the deep quantum regime, Eq.~\eqref{eq:qhigh_lambda_deep}, plus higher-order corrections which scale with powers of the quantum parameter $\alpha_N$. Since we demand $\alpha_N\ll 1$, we can neglect these higher-order corrections in the deep quantum regime, which completes our proof to justify the two-level approximation, at least in the exponential-gain limit. 

In addition, the expression in Eq.~\eqref{eq:qhigh_lambda_corr} enables us to compare our result  with existing FEL literature. In Ref.~\cite{boni06}, for example, the cubic equation   
\ba\label{eq:qhigh_cubic_boni}
 \left(\lambda^2-1\right)\left(\lambda +1 +\varkappa\alpha_N\right)-2\alpha_N^2=0\,,
\ea
which is written in the scaling of the present article, was derived. 

The approach in Ref.~\cite{boni06} leading to Eq.~\eqref{eq:qhigh_cubic_boni} was based on the introduction of a collective bunching operator and a -- properly ordered -- momentum bunching operator in analogy to classical FEL theory~\cite{bnp}. The corresponding equations of motion were first linearized and solved by the ansatz $\e{-\I\lambda\tau}$. From Eq.~\eqref{eq:qhigh_cubic_boni} it was then possible to asymptotically find the correct classical limit by setting $\alpha_N\gg 1$. The opposite case, that is $\alpha_N\ll 1$, was then identified as the quantum regime. 

We emphasize that our approach takes the opposite route: we first searched for a quantum regime by identifying  the two important time scales directly in the Hamiltonian, Eq.~\eqref{eq:qhigh_HI}, and reduced it in lowest order of $\alpha_N$ to the Dicke Hamiltonian. We then solved the simplified  equations of motion and obtained the analytic result, Eq.~\eqref{eq:qhigh_lambda_deep}, for the growth rate of the laser intensity, as well as the expression including higher-order corrections, Eq.~\eqref{eq:qhigh_lambda_corr}.    

\begin{figure}
\centering
\includegraphics[width=0.45 \textwidth]{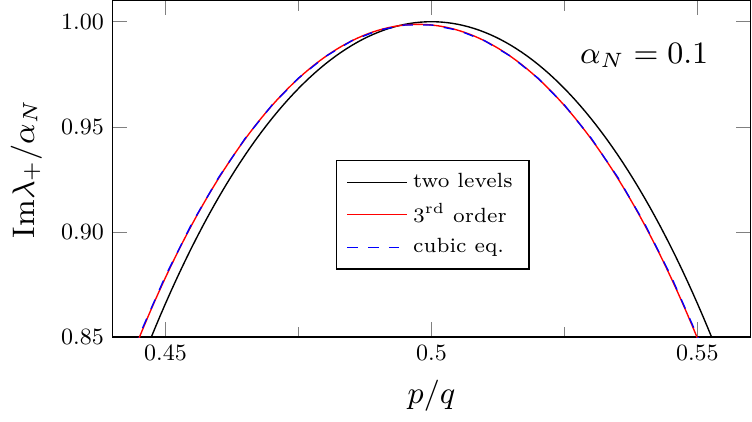}
\includegraphics[width=0.45 \textwidth]{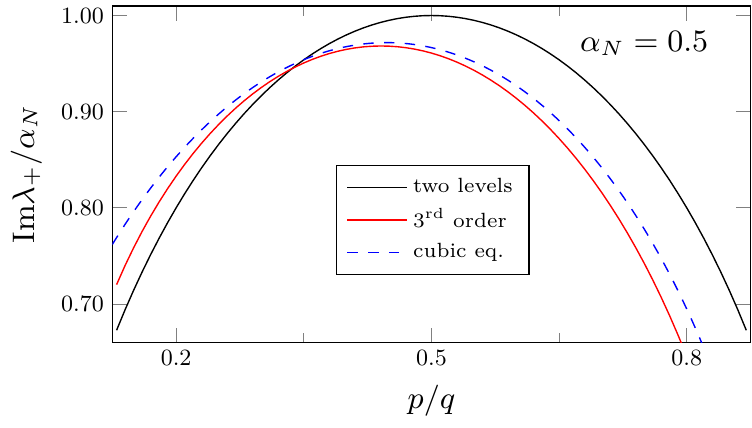}
\caption{Gain function of a high-gain Quantum FEL including higher-order corrections: we have drawn the increment $\text{Im}\lambda_+$ for the growth of the mean photon number against the initial momentum $p$ of the electrons divided by the recoil $q$ for two different values of the quantum parameter, that is $\alpha_N=0.1$ (top) and $\alpha_N=0.5$ (bottom). In both plots, we have compared the first-order solution, Eq.~\eqref{eq:qhigh_lambda_deep}, (black line) corresponding to the two-level approximation and the expression, Eq.~\eqref{eq:qhigh_lambda_corr}, (red line) in third order to the numerical solution of the cubic equation, Eq.~\eqref{eq:qhigh_cubic_boni}, (blue dashed line) from Ref.~\cite{boni06}. In the deep quantum regime, $\alpha_N=0.1$, we observe that all three curves approximately agree with a nearly perfect matching of the third-order solution, Eq.~\eqref{eq:qhigh_lambda_corr}, with the result of Ref.~\cite{boni06}. If we increase the quantum parameter to $\alpha_N=0.5$, the two-level approximation significantly differs from the other two curves. However, the third-order solution and the result from Ref.~\cite{boni06} still agree very well despite a small shift that originates from even higher orders of $\alpha_N$. Moreover, while the two-level approximation leads to a maximized gain at $p=q/2$,  the maximum in third order has moved to the left  for increasing values of $\alpha_N$. This behavior is consistent with the transition of the quantum resonance at $p=q/2$ to the classical one, $p=0$, in Fig.~\ref{fig:high_cl_q}.}
\label{fig:high_deep_corr_boni}
\end{figure}

In Fig.~\ref{fig:high_deep_corr_boni} we have drawn the the gain function of a Quantum FEL, that is $\text{Im}\lambda_+$ depending on $p$, for $\alpha_N=0.1$ (top) and $\alpha_N=0.5$ (bottom), respectively. 
Moreover, we compare our results, that is  Eq.~\eqref{eq:qhigh_lambda_deep} from the two-level approximation, as well as the third-order expression, Eq.~\eqref{eq:qhigh_lambda_corr}, to the numerical solution of the cubic equation, Eq.~\eqref{eq:qhigh_cubic_boni}, from Ref.~\cite{boni06}.

We make two important observations from Fig.~\ref{fig:high_deep_corr_boni}: (i) The maximum of the gain function for the deep quantum regime always occurs at $p=q/2$. In contrast, the maximum of the higher-order result is located  slightly on the left of $q/2$. This shift already hints the transition from the quantum resonance $p=q/2$ to the classical one $p=0$ apparent in Fig.~\ref{fig:high_cl_q}. We note that an analogous effect occurs in the field of atomic diffraction, where it is known as `light shift'~\cite{giese16}.     

(ii) In the deep quantum regime, exemplified by $\alpha_N=0.1$, we observe three very similar curves with an almost perfect agreement between our third-order solution  and the result of Ref.~\cite{boni06}. Increasing the quantum parameter to $\alpha_N=0.5$, however, leads to a growing deviation of the latter two curves to  Eq.~\eqref{eq:qhigh_lambda_deep} describing the two-level approximation. This result is not surprising since we are outside the deep quantum regime and we do not expect that the two-level approximation gives us here a perfectly correct description of the FEL dynamics. The higher-order result as well as the solution~\cite{boni06} of the cubic equation, however, still show a good agreement apart from a small deviation. We interpret this small shift as an effect from higher orders in $\alpha_N$ than the third one. 

We deduce from Fig.~\ref{fig:high_deep_corr_boni} that the two different approaches, the method of averaging of the present article and the procedure in Ref.~\cite{boni06}, lead to equivalent results~\footnote{This equivalence which we have established via graphical means in Fig.~\ref{fig:high_deep_corr_boni} can be also proven in a rigorous, analytic manner. Therefore, we expand the 
analytic solution~\cite{bronstein} of the cubic equation, Eq.~(\ref{eq:qhigh_cubic_boni}), in powers of $\alpha_N$ or alternatively 
solve it asymptotically with an iteration method~\cite{peter,hinch}.} 
%
%
%
in the asymptotic limit $\alpha_N\ll 1$.

%% file: Conclusions.tex
\section{Conclusions}
\label{sec:Conclusions}

In this article we have used collective jump operators and a rotating wave-like approximation to establish the analogy of a high-gain Quantum FEL to the Dicke model of standard quantum optics. By performing a parametric approximation, we have derived analytic expressions for the mean photon number,  the gain length, the gain function, and the gain bandwidth in this regime. Moreover, we have obtained a super-Poissonian photon statistics of the emitted radiation.

With the help of the method of canonical averaging we have proved the two-level behavior of the electron dynamics in a rigorous manner. In this context, the quantum parameter $\alpha_N$ occurs as the expansion
parameter of the corresponding asymptotic series. We, moreover, have embedded our approach into a broader context by showing that our results are consistent with  existing literature~\cite{boni06} on the Quantum FEL. In contrast to these earlier approaches, we identify the two-level limit of the FEL dynamics directly in the Hamiltonian with the benefit of simple analytic results.     

In an upcoming article we plan to extend the present theory by leaving the exponential-gain regime and investigating the long-time behavior of the dynamics. Moreover, we will discuss the experimental requirements for a high-gain Quantum FEL based on the predictions of our intuitive approach and on the detailed study in Ref.~\cite{debus2018}.   

Although our elementary model explains the fundamental effects of a Quantum FEL and provides us with important scaling laws it is still not a \emph{complete} theory of such a device. Therefore, one has to consider at least two additional effects. To understand which parts of the electron beam really
communicate with each other through the laser field one has to take the slippage of the radiation pulse over the electrons~\cite{boni_sase} into account. In addition, decoherence effects like space charge~\cite{sprangle1} or spontaneous emission into all modes~\cite{robb_se2012} can negatively affect the Quantum FEL dynamics~\cite{debus2018}.    
  
Indeed, today's technological possibilities prevent the operation of a low-gain Quantum FEL oscillator and we have to consider the high-gain regime  for an experimental realization.
However, first concepts~\cite{xfelo1,*xfelo2} emerged to construct high-quality mirrors in the X-ray regime and to operate X-ray FEL oscillators. The resulting implications for a possible Quantum FEL oscillator  will be discussed elsewhere.

%% file: Transformation_of_Hamiltonian_in_terms_of_jump_operator.tex
\section{Transformation of Hamiltonian in terms of jump operators}
\label{app:Transformation_of_Hamiltonian_in_terms_of_jump_operator}

In this appendix we derive the transformed Hamiltonian, Eq.~\eqref{eq:qhigh_HI}, which we employ for our analysis of the quantum regime, from the original Hamiltonian, Eq.~\eqref{eq:H_many}, in the Heisenberg picture. Therefore, we first express the occurring operators in terms of the collective jump operators $\hat{\Upsilon}_{\mu,\nu}$. In the next step we perform the transition into a rotating frame analogous to the interaction picture.

\subsection{Expressing terms with jump operators}

The Hamiltonian, Eq.~\eqref{eq:H_many}, consists of sums of single-electron operators $\hat{\mathcal{O}}_j$ which can be written as 
\ba 
\sum\limits_{j=1}^N \hat{\mathcal{O}}_j=\sum\limits_{j=1}^N\sum\limits_{\mu,\nu} 
{}_{(j)}\bra{p-\mu q}\hat{\mathcal{O}}_j\ket{p-\nu q}_{(j)} \sigma_{\mu,\nu}^{(j)}\,,
\ea
where we have assumed that the infinite momentum ladder for one electron constitutes a complete set of basis states and have introduced the single-electron jump operators $\hat{\sigma}_{\mu,\nu}^{(j)}$ from Eq.~\eqref{eq:sigmamunu_def}.

Because all electrons shall possess the same initial state, that is $\ket{p,p,...,p}$, and with the help of the definition, Eq.~\eqref{eq:qhigh_proh}, for the collective operators $\displaystyle\hat{\Upsilon}_{\mu,\nu}$ we straightforwardly derive the identities 
\ba\label{eq:app_qhigh_terms_coll} 
\begin{cases}
\sum\limits_{j=1}^N\hat{p}_j^2=\sum\limits_\mu\left(p-\mu q\right)^2\hat{\Upsilon}_{\mu,\mu}\\
\sum\limits_{j=1}^N\e{\I2k\hat{z}_j}=\sum\limits_{\mu}\hat{\Upsilon}_{\mu,\mu+1}\\
\sum\limits_{j=1}^N\e{-\I2k\hat{z}_j}=\sum\limits_{\mu}\hat{\Upsilon}_{\mu+1,\mu}
\end{cases}
\ea
for each term of the Hamiltonian, Eq.~\eqref{eq:H_many}. 

In terms of the jump operators the total  Hamiltonian 
\ba
\hat{H}\equiv\hat{H}_0+\hat{H}_1
\ea
is given by 
\ba\label{eq:app_qhigh_Hfree} 
\hat{H}_0 = \sum\limits_\mu \left(\del+\frac{1}{2}-\mu\right)^2\hat{\Upsilon}_{\mu,\mu}
\ea 
which denotes the free motion of the electrons, and by  
\ba\label{eq:app_qhigh_Hww} 
\hat{H}_1\equiv \varepsilon \left(\hat{a}_\L \sum\limits_\mu \hat{\Upsilon}_{\mu,\mu+1}+\hat{a}_\L^ \dagger\sum\limits_\mu \hat{\Upsilon}_{\mu+1,\mu}\right)
\ea
which describes the interaction of the electrons with the laser field. Here we have introduced the
dimensionless coupling $\varepsilon \equiv g/\omega_\text{r}$ with the recoil frequency $\omega_\text{r}$, Eq.~\eqref{eq:rec_freq}, and the relative deviation 
$\del$, Eq.~\eqref{eq:delta_def}, of the momentum $p$ from $p=q/2$.

The dynamics of an operator $\hat{\mathcal{O}}$ in the Heisenberg picture is dictated by the equation of motion 
\ba\label{eq:hberg_eqmot_orig} 
\I\frac{\D}{\D\taub}\hat{\mathcal{O}}(\taub)
=\left[\hat{\mathcal{O}}(\taub),\hat{H}_0\right]+\left[\hat{\mathcal{O}}(\taub),\hat{H}_1\right]\,,
\ea
where the time $\taub\equiv \omega_\text{r}t$ is written in a dimensionless  form.

\subsection{Transformation in rotating frame}

Similarly to the transformation to the interaction picture, we move to a frame, where the dynamics corresponding to the free motion, that is $\hat{H}_0$, is removed from the total Hamiltonian, but its effect is accounted for by time-dependent phases. Moreover, we try to avoid that any contribution including the deviation $\del$ from $q/2$ appears in the phases of the Hamiltonian, since it corresponds to a slowly-varying dynamics, that is $\del\ll 1$. 

Hence, we perform the transformation   
\ba\label{eq:app_qhigh_trafo2} 
 \begin{cases}
\hat{\mathcal{O}}'(\taub)\equiv \e{-\I\taub\left(\hat{H}_0+\del\hat{n}\right)}\hat{\mathcal{O}}\e{\I\taub\left(\hat{H}_0+\del\hat{n}\right)}\\
\hat{H}'(\taub)\equiv \e{-\I\taub\left(\hat{H}_0+\del\hat{n}\right)}\hat{H}_1\e{\I\taub\left(\hat{H}_0+\del\hat{n}\right)}\\
\ket{\Psi'(\taub)}\equiv \e{-\I\taub\left(\hat{H}_0+\del\hat{n}\right)}\ket{\Psi(\taub)}
 \end{cases}
\ea
from the Heisenberg picture to the rotating frame.
This procedure finally leads with the help of Eq.~\eqref{eq:hberg_eqmot_orig} and the commutation relation Eq.~\eqref{eq:qhigh_comm} to the transformed Hamiltonian
\begin{equation}\label{eq:app_qhigh_HI}
\hat{H}'(\taub)=\varepsilon  \left(\hat{a}_\L \sum\limits_\mu \e{\I2\mu\taub} \hat{\Upsilon}_{\mu,\mu+1}
+\text{h.c.}\right) 
-\del \, \hat{n}
\end{equation}
which, together with the equation of motion
\ba\label{eq:app_eqofmot} 
\I\frac{\D}{\D\taub}\hat{\mathcal{O}}'(\taub)=\left[\hat{\mathcal{O}}'(\taub),\hat{H}'(\taub)\right]
\ea
forms the basis of our approach towards the high-gain Quantum FEL. 

We emphasize that this transformation is only useful, if all contributions of $\hat{H}'$ are of comparable order of magnitude, that is $\del\sim \varepsilon$ and $\hat{n}\equiv \hat{a}_\L^\dagger\hat{a}_\L$ initially is small. Hence, our discussion is restricted to a seeded FEL with a small seeding or to the SASE mode of operation.

%% file: Canonical_averaging.tex
\section{Canonical averaging}
\label{app:Canonical_averaging}

In this appendix we present the asymptotic method of canonical averaging~\cite{bogoliubov,higher} which we use to derive the linearized equations of motions in the deep quantum regime as well as for higher orders, that is Eqs.~\eqref{eq:qhigh_lambda_deep} and ~\eqref{eq:qhigh_lambda_corr}, respectively. For this purpose, we begin with a general description of the method, before we apply it on the FEL and finally linearize the resulting equations of motion.

\subsection{Description of method}

The method of averaging is based on the  distinction between slowly-varying and rapidly-varying contributions of the dynamics. Hence, we first perform a transformation which separates these two parts and derive a formal expression for an effective Hamiltonian corresponding to the slowly-varying dynamics. The explicit form of this Hamiltonian is then found by a perturbative expansion with the constraint that
the effective Hamiltonian is in each order independent of time. By this procedure we avoid  secularly-growing terms which would occur in ordinary perturbation theory. We explicitly perform this asymptotic expansion up to terms in third order.    

\subsubsection{Slowly- and rapidly-varying terms}

We start by considering a Hamiltonian which can be written as the Fourier series
\ba\label{eq:app_Hfour} 
\hat{H}'(\taub)=\varepsilon \sum\limits_\mu \hat{\mathcal{H}}_\mu \e{\I 2\mu\taub}
\ea
with $\varepsilon\ll 1$. This Hamiltonian leads to rapid oscillations with multiples of $\tau$. However, due to the term with $\mu=0$ or possible  cross terms, where the time-dependent phases cancel, also slowly-varying contributions to the dynamics emerge. 

Therefore, we assume that the time evolution of an operator  
\ba 
\hat{\mathcal{O}}'(\taub)\equiv \e{-\hat{F}(\taub)}\hat{\chi}(\taub)\e{\hat{F}(\taub)}
\ea
can be separated into a slowly-varying  and a rapidly-varying part, $\hat{\chi}$ and $\hat{F}$, respectively. In  Ref.~\cite{higher} an analogous procedure was carried out for the density operator $\hat{\rho}$.  

From the Heisenberg equation, Eq.~\eqref{eq:hber_eqmot_ht}, for $\hat{\mathcal{O}}'$ we derive the transformed equation of motion
\ba 
\I \frac{\D}{\D\taub}\hat{\chi}(\taub)=\left[\hat{\chi}(\taub),\hat{H}_\text{eff}\right]
\ea
for the slowly-varying part $\hat{\chi}$ with the effective Hamiltonian
\ba\label{eq:Heff_formal}
\hat{H}_\text{eff}=\e{\hat{F}(\taub)}\hat{H}'(\taub)\e{-\hat{F}(\taub)}-\I\,\frac{\text{d}\e{\hat{F}(\taub)}}{\text{d}\taub}\e{-\hat{F}(\taub)}\,.
\ea
In the course of this derivation we have made use of the relation
\ba 
\frac{\text{d}\e{-\hat{F}(\tau)}}{\text{d}\taub}=-\e{-\hat{F}(\tau)}\frac{\text{d}\e{\hat{F}(\tau)}}{\text{d}\taub}\e{-\hat{F}(\tau)}\,,
\ea
which  straightforwardly follows from the derivative of an inverse operator.

With the help of the Baker--Campbell--Hausdorff formula~\cite{louisell} we write the first term in Eq.~\eqref{eq:Heff_formal} as
\ba\label{eq:Heff_formal_eins} 
\e{\hat{F}(\taub)}\hat{H}'(\taub)\e{-\hat{F}(\taub)}=\sum\limits_{j=0}^\infty \frac{1}{j!}\left[\hat{F}(\taub),\hat{H}'(\taub)\right]_j\,,
\ea
where the nested commutators
\ba 
[\hat{A},\hat{B}]_{j}\equiv &[\hat{A},[\hat{A},\hat{B}]_{j-1}]
\ea
for $j\neq 0$ and
\ba
[\hat{A},\hat{B}]_{0}\equiv &\hat{B}
\ea
are defined in a recursive way. 

According to Ref.~\cite{higher} we cast the second term of Eq.~\eqref{eq:Heff_formal} into the form
\begin{equation}\label{eq:Heff_formal_zwei}
\frac{\text{d}\e{\hat{F}(\taub)}}{\text{d}\taub}\e{-\hat{F}(\taub)}=\sum\limits_{j=0}^\infty\frac{1}{(j+1)!}\left[\hat{F}(\taub),\frac{\text{d}\hat{F}(\taub)}{\text{d}\taub}\right]_j\,.
\end{equation}
An elegant proof of this identity can be found in Ref.~\cite{rossmann}.
We emphasize that we have not done any approximation up to this point and the expressions in Eqs.~\eqref{eq:Heff_formal},~\eqref{eq:Heff_formal_eins} and~\eqref{eq:Heff_formal_zwei} are still exact.

\subsubsection{Perturbative expansion \& avoiding secular terms}

When we assume that $\varepsilon |\hat{\mathcal{H}}_\mu|\ll 1$~\cite{higher} we are allowed to perform perturbative expansions for the rapidly varying terms, that is
\ba\label{eq:app_coav_F_exp}
\hat{F}(\taub)=\varepsilon \hat{F}^{(1)}(\taub) +\varepsilon^2 \hat{F}^{(2)}(\taub) +\varepsilon^3 \hat{F}^{(3)}(\taub)+...
\ea
and for the effective Hamiltonian, that is
\ba\label{eq:app_coav_Heff_exp} 
\hat{H}_\text{eff}=\varepsilon \hat{H}_\text{eff}^{(1)} +\varepsilon^2 \hat{H}_\text{eff}^{(2)} +\varepsilon^3 \hat{H}_\text{eff}^{(3)}+...\,,
\ea
both in powers of $\varepsilon$.

The main difference of the method of averaging to ordinary perturbation theory is given by the constraint
\ba\label{eq:app_avoid_secular} 
\hat{H}_\text{eff}^{(k)}\neq \hat{H}_\text{eff}^{(k)}(\taub)
\ea
which means 
that the effective Hamiltonian  does not contain any time-dependent term, but includes all time-independent terms in each order of the expansion. In contrast, if $\hat{F}$ had any time-independent contribution, as it is the case in ordinary perturbation theory, we would observe secularly-growing terms which are unphysical. Moreover, we note that the slowly-varying dynamics, dictated by the effective Hamiltonian, has to be solved in a non-perturbative way -- else we would have gained nothing in comparison to standard perturbation theory.

\subsubsection{First order}

Inserting the expression, Eq.~\eqref{eq:app_Hfour}, for the full Hamiltonian into the relation, Eq.~\eqref{eq:Heff_formal} together with Eqs.~\eqref{eq:Heff_formal_eins} and~\eqref{eq:Heff_formal_zwei}, for the effective one  and keeping only first-order terms of the expansions, Eqs.~\eqref{eq:app_coav_F_exp} and~\eqref{eq:app_coav_Heff_exp}, yields  the relation
\ba\label{eq:app_Heff1_zw} 
\hat{H}_\text{eff}^{(1)}= \hat{\mathcal{H}}_0+\underbrace{\left(\sum\limits_{\mu \neq 0}\hat{\mathcal{H}}_\mu\e{\I 2\mu\taub}-\I\,\frac{\D \hat{F}^{(1)}(\taub)}{\D\taub}
\right)}_{=0}\,.
\ea
In order to satisfy the constraint, Eq.~\eqref{eq:app_avoid_secular}, we identify  the first term in Eq.~\eqref{eq:app_Heff1_zw} as the  lowest-order contribution  
\ba\label{eq:app_coav_Heff1} 
\hat{H}_\text{eff}^{(1)}=\hat{\mathcal{H}}_0
\ea
to the effective Hamiltonian, which is simply the time-independent part of the full Hamiltonian, Eq.~\eqref{eq:app_Hfour}. 

Moreover, we require that the time-dependent terms in parentheses vanish which means that they have to be absorbed in $\hat{F}^{(1)}$. Hence, we obtain
\ba\label{eq:app_coav_F1} 
\hat{F}^{(1)}(\taub)=-\sum\limits_{\mu\neq 0}\frac{\e{\I 2\mu\taub}}{2\mu}\hat{\mathcal{H}}_\mu\,,
\ea
where we have integrated over time $\taub$.

\subsubsection{Second order}

From Eqs.~\eqref{eq:Heff_formal},~\eqref{eq:Heff_formal_eins} and~\eqref{eq:Heff_formal_zwei}  as well as from the expansions, Eqs.~\eqref{eq:app_coav_F_exp} and~\eqref{eq:app_coav_Heff_exp}, we find the expression
\begin{equation}\label{eq:app_Heff2_zw} 
 \begin{aligned}
\hat{H}_\text{eff}^{(2)}= \left[\hat{F}^{(1)}(\taub),\sum\limits_{\mu=0}\hat{\mathcal{H}}_\mu\e{\I 2\mu\taub}\right]
-\I\, \frac{\D\hat{F}^{(2)}(\taub)}{\D\taub} \\
-\frac{1}{2}\left[\hat{F}^{(1)}(\taub),\I\,\frac{\D\hat{F}^{(1)}(\taub)}{\D\taub}\right]\,,
 \end{aligned}
\end{equation}
where we have considered only terms which are quadratic in $\varepsilon$.

By inserting $\hat{F}^{(1)}$ from Eq.~\eqref{eq:app_coav_F1} into Eq.~\eqref{eq:app_Heff2_zw}  and applying the prescription in Eq.~\eqref{eq:app_avoid_secular} we straightforwardly derive the second-order contributions 
\ba\label{eq:app_coav_Heff2} 
\hat{H}_\text{eff}^{(2)}=-\frac{1}{2}\sum\limits_{\nu\neq 0}\frac{1}{2\nu}
\left[\hat{\mathcal{H}}_\nu,\hat{\mathcal{H}}_{-\nu}\right]
\ea
and
\ba\label{eq:app_coav_F2}
\hat{F}^{(2)}(\taub)=\frac{1}{2}\!\sum\limits_{\substack{\mu,\rho \neq 0\\ \mu \neq \rho}}\!
\frac{\e{\I 2\rho \tau}}{4\mu\rho}
\left[\hat{\mathcal{H}}_\mu,\hat{\mathcal{H}}_{\rho-\mu}\right]
+\sum\limits_{\mu\neq 0} \frac{\e{2\I\mu\taub}}{4\mu^2}
\left[\hat{\mathcal{H}}_\mu,\hat{\mathcal{H}}_0\right]
\ea
to the effective Hamiltonian and the rapidly-varying dynamics, respectively.

\subsubsection{Third order}

In third order of $\varepsilon$ we obtain 
\begin{widetext}
\ba \label{eq:app_coav_3}
\hat{H}_\text{eff}^{(3)}=
\left[\hat{F}^{(2)}(\taub),\sum\limits_{\mu=0}\hat{\mathcal{H}}_\mu\e{2\I\mu\taub}\right]
+\frac{1}{2}\left[\hat{F}^{(1)}(\taub),\left[\hat{F}^{(1)}(\taub), 
\sum\limits_{\mu=0}\hat{\mathcal{H}}_\mu\e{\I2\mu\tau}\right]\right]
-\I\, \frac{\D\hat{F}^{(3)}(\taub)}{\D\taub}
-\frac{1}{2}\left[\hat{F}^{(1)}(\taub),\I\,\frac{\D\hat{F}^{(2)}(\taub)}{\D\taub}\right]\\
-\frac{1}{2}\left[\hat{F}^{(2)}(\taub),\I\,\frac{\D\hat{F}^{(1)}(\taub)}{\D\taub}\right]
-\frac{1}{6}\left[\hat{F}^{(1)}(\taub),\left[\hat{F}^{(1)}(\taub),\I\,\frac{\D\hat{F}^{(1)}(\taub)}{\D\taub}\right]\right]
\ea
\end{widetext}
which leads with the help of Eqs.~\eqref{eq:app_coav_F1} and~\eqref{eq:app_coav_F2} as well as with the condition Eq.~\eqref{eq:app_avoid_secular} to the expression
\begin{equation}\label{eq:app_coav_Heff3}
 \begin{aligned} 
\hat{H}_\text{eff}^{(3)}=-\frac{1}{3}\sum\limits_{\substack{\mu,\rho \neq 0\\ \mu+\rho\neq 0}}\frac{1}{4\mu(\mu+\rho)}\left[\hat{\mathcal{H}}_{-(\mu+\rho)},
\left[\hat{\mathcal{H}}_{\mu},\hat{\mathcal{H}}_\rho\right]\right]\\
-\frac{1}{2}\sum\limits_{\mu\neq 0}\frac{1}{4\mu^2}
\left[\hat{\mathcal{H}}_\mu,\left[\hat{\mathcal{H}}_{-\mu},\hat{\mathcal{H}}_0\right]\right]\,,
 \end{aligned}
\end{equation}
for the effective Hamiltonian.

\subsection{Application to the Quantum FEL}

We now apply the equations from the  method of averaging, which we have derived in the previous section, on the FEL Hamiltonian, Eq.~\eqref{eq:qhigh_Hfour}. After that, we linearize the resulting equations of motion which enables us to obtain higher-order corrections to the gain of the deep quantum regime, Eq.~\eqref{eq:qhigh_lambda_deep}. 

\subsubsection{Effective Hamiltonian}

The Fourier components of the FEL Hamiltonian, Eq.~\eqref{eq:qhigh_Hfour}, are given by, Eq.~\eqref{eq:qhigh_Fcomp},   
\ba\label{eq:app_many_Fcomp}
 \begin{cases}
  \hat{\mathcal{H}}_0 &= \hat{a}_\text{L}\hat{\Upsilon}_{0,1} + \hat{a}^{\dagger}_\text{L}\hat{\Upsilon}_{1,0} -\frac{\del}{\varepsilon}\;\hat{n}\\
  \hat{\mathcal{H}}_\mu & = \hat{a}_\text{L} \hat{\Upsilon}_{\mu,\mu+1} + \hat{a}^{\dagger}_\text{L} \hat{\Upsilon}_{-\mu+1,-\mu}\,.
 \end{cases}
\ea
By inserting these components into Eqs.~\eqref{eq:app_coav_Heff1} and~\eqref{eq:app_coav_Heff2} and by employing the commutation relations for the jump operators, Eq.~\eqref{eq:qhigh_comm}, and for the laser-field operators 
we obtain 
\ba\label{eq:app_many_H1}
\hat{H}^{(1)}_\text{eff}=\hat{a}_\text{L}\hat{\Upsilon}_{0,1} + \hat{a}^\dagger_\text{L}\hat{\Upsilon}_{1,0}-\frac{\del}{\varepsilon}\;\hat{n}
\ea
and
\ba\label{eq:app_many_H2}
\hat{H}^{(2)}_\text{eff}= \frac{1}{2}\left(\hat{n}+1\right)
\sum\limits_{\mu\neq 0}\frac{1}{\mu}\left(\hat{\Upsilon}_{\mu+1,\mu+1}-\hat{\Upsilon}_{\mu,\mu}\right)
\\- \sum\limits_{\mu\neq 0}\frac{1}{\mu}
\hat{\Upsilon}_{\mu+1,\mu}\hat{\Upsilon}_{\mu,\mu+1}
\ea
for the effective Hamiltonian in first and second order of the method of averaging, respectively. 

An analogous procedure leads with the help of Eq.~\eqref{eq:app_coav_Heff3} to the effective Hamiltonian
\ba\label{eq:app_many_H3}
\hat{H}_\text{eff}^{(3)}=\hat{H}_\text{lin}^{(3)}+\hat{H}_\text{cub}^{(3)}+\hat{H}_\del^{(3)}
\ea
in third order, where 
\begin{widetext}
\ba 
\hat{H}_\text{lin}^{(3)}=\frac{\hat{a}_\L}{4}\!\left[\!\sum\limits_{\substack{\mu\neq 0 \\ \mu\neq -1}}\!\!\frac{\hat{\Upsilon}_{2\mu+2,2\mu+1}\hat{\Upsilon}_{\mu,\mu+2}}{\mu (\mu+1)(2\mu+1)}
 -\frac{1}{2} \sum\limits_{\mu \neq 0} 
\frac{1}{\mu^2}\!\left(\hat{\Upsilon}_{\mu+1,\mu+1}-\hat{\Upsilon}_{\mu,\mu}\right)\hat{\Upsilon}_{0,1}+\frac{3}{2}\left(\hat{\Upsilon}_{0,-1}\hat{\Upsilon}_{-1,1}
-\hat{\Upsilon}_{0,2}\hat{\Upsilon}_{2,1}\right)\!+\frac{1}{2}\hat{\Upsilon}_{0,1}\right]\!+\!\text{h.c.}
\ea
\end{widetext}
denotes the term linear in the field operators $\hat{a}_\L$ and $\hat{a}_\L^ \dagger$, while  
\ba 
\hat{H}_\text{cub}^{(3)}=\frac{1}{4}\hat{a}_\L^3\hat{\Upsilon}_{-1,2}
-\frac{1}{4}\left(\hat{a}_\L^\dagger\hat{a}_\L^2 +\hat{a}_\L^2\hat{a}_\L^\dagger \right)\hat{\Upsilon}_{0,1} +\text{h.c.}
\ea
contains cubic combinations of the field operators and 
\ba 
\hat{H}_\del^{(3)}=
\frac{\del}{4\varepsilon}\left[(\hat{n}+1) \sum\limits_{\mu\neq 0}\frac{1}{\mu^2} 
\left(\hat{\Upsilon}_{\mu+1,\mu+1}-\hat{\Upsilon}_{\mu,\mu}\right)\right.\\
\left.-\sum\limits_{\mu\neq 0}\frac{1}{\mu^2}\hat{\Upsilon}_{\mu+1,\mu}\hat{\Upsilon}_{\mu,\mu+1} \right]
\ea
is the contribution arising from a nonzero deviation $\del$ from resonance $p=q/2$.

\subsubsection{Linearization procedure}

In order to obtain the time evolution of an operator $\hat{\mathcal{O}}'\cong \hat{\chi}$ we have to solve the Heisenberg equation of motion
\begin{equation}\label{eq:app_coav_eqmotthird} 
\I\frac{\D}{\D\taub}\hat{O}'\cong\varepsilon \left[\hat{\mathcal{O}}',\hat{H}_\text{eff}^{(1)}\right]+\varepsilon^2 \left[\hat{\mathcal{O}}',\hat{H}_\text{eff}^{(2)}\right]
+\varepsilon^3 \left[\hat{\mathcal{O}}',\hat{H}_\text{eff}^{(3)}\right]
\end{equation}
with the effective Hamiltonian given in Eqs.~\eqref{eq:app_many_H1},~\eqref{eq:app_many_H2} and~\eqref{eq:app_many_H3}. 
Unfortunately, Eq.~\eqref{eq:app_coav_eqmotthird} corresponds to a set of non-linearly coupled differential equations for non-commuting operators. To find an analytic solution we linearize Eq.~\eqref{eq:app_coav_eqmotthird} in analogy to the parametric approximation~\cite{kumar} for the Dicke Hamiltonian. 

In more detail, we assume that initially all electrons populate the same level $p\sim q/2$  and that for short times only a few electrons jump to different levels. Hence, we can replace the operator $\hat{\Upsilon}_{0,0}$ by its expectation value at $\taub=0$, that is $\hat{\Upsilon}_{0,0}\cong N\gg 1$. 
In contrast, we treat $\hat{a}_\L\cong \updelta \hat{a}_\L$ and $\hat{\Upsilon}_{\mu,\nu}\cong\updelta\hat{\Upsilon}_{\mu,\nu}$ (except for $\hat{\Upsilon}_{0,0}$) as small quantities. Thus, we only keep contributions linear in these operators, while we discard quadratic or higher-order combinations of them. 
We emphasize, however, that the validity of this procedure is restricted to comparatively short interaction times, where we can treat $\hat{\Upsilon}_{0,0}$ as constant.

In first order of the method of averaging we derive the linearized commutators
\ba\label{eq:app_many_Y1}
\left[\hat{\Upsilon}_{1,0},\hat{H}^{(1)}_\text{eff}\right]\cong -N\hat{a}_\text{L}
\ea
and
\ba\label{eq:app_many_a1}
\left[\hat{a}_\text{L},\hat{H}^{(1)}_\text{eff}\right]=\hat{\Upsilon}_{1,0}-\frac{\del}{\varepsilon}\hat{a}_\text{L}
\ea
for $\hat{\Upsilon}_{1,0}$ and $\hat{a}_\L$, respectively, with $\hat{H}_\text{eff}^{(1)}$ from  Eq.~\eqref{eq:app_many_H1}, where we have employed the commutation relation, Eq.~\eqref{eq:qhigh_comm} for the jump operators.

Analogous procedures yield the relations 
\ba\label{eq:app_many_Y2}
\left[\hat{\Upsilon}_{1,0},\hat{H}^{(2)}_\text{eff}\right]
\cong -\frac{1}{2}\hat{\Upsilon}_{1,0}
\ea
and
\ba\label{eq:app_many_a2}
\left[\hat{a}_\text{L},\hat{H}^{(2)}_\text{eff}\right]\cong -\frac{1}{2} N \hat{a}_\text{L}\,,
\ea
for second order as well as
\ba\label{eq:app_many_Y3}
\left[\hat{\Upsilon}_{1,0},\hat{H}^{(3)}_\text{eff}\right]
\cong -\frac{N^2}{8}\hat{a}_\L +\frac{\del}{4\varepsilon}\hat{\Upsilon}_{1,0}
\ea
and
\ba\label{eq:app_many_a3}
\left[\hat{a}_\text{L},\hat{H}^{(3)}_\text{eff}\right]\cong \frac{N}{8}\hat{\Upsilon}_{1,0}-\frac{\del}{4\varepsilon}\hat{a}_\L
\ea
for third order, where we have used Eqs.~\eqref{eq:app_many_H2} and~\eqref{eq:app_many_H3}, respectively.

After rescaling $\hat{\Upsilon}_{1,0}$ via the prescription
\ba
\hat{\mathcal{Y}}_{1,0}&=\frac{1}{\sqrt{N}}\hat{\Upsilon}_{1,0}\\
\ea
in analogy to Eq.~\eqref{eq:qhigh_tilde}  and inserting the relations Eqs.~\eqref{eq:app_many_Y1} to~\eqref{eq:app_many_a3}  into Eq.~\eqref{eq:app_coav_eqmotthird} we obtain the linearized set of equations
\ba\label{eq:app_lin_corr}
\I\frac{\text{d}}{\text{d}\tau}
 \begin{pmatrix}
 \updelta \hat{\mathcal{Y}}_{1,0} \\ \updelta\hat{a}_\text{L}
 \end{pmatrix}=M \begin{pmatrix}
  \updelta\hat{\mathcal{Y}}_{1,0} \\ \updelta\hat{a}_\text{L}
 \end{pmatrix}
\ea
for  $\hat{a}_\L\cong \updelta \hat{a}_\L$ and $\hat{\mathcal{Y}}_{1,0}\cong\updelta\hat{\mathcal{Y}}_{1,0}$, where the matrix
\ba 
M\equiv\begin{pmatrix}
 0 & -\alpha_N\left(1-\frac{\alpha_N^2}{8}\right)\\
\alpha_N\left(1-\frac{\alpha_N^2}{8}\right) & -\alpha_N \left(\varkappa +\frac{\alpha_N}{2} -\frac{\varkappa\alpha_N^2}{4}\right)
\end{pmatrix}
\ea
includes contributions up to third order in the quantum parameter $\alpha_N\equiv \varepsilon\sqrt{N}$.

%% file: Kling_FEL.bbl
\begin{thebibliography}{85}
\bibitem [{fra()}]{frame}%
  \BibitemOpen
  \href@noop {} {}\bibinfo {note} {{S}trictly speaking the Bambini--Renieri
  frame is only defined if there exists a well defined laser mode which is the
  case for a seeded FEL. For SASE, this assumption is therefore only an
  approximation. However, in the quantum regime all electrons have almost the
  same momentum and thus emit into similar laser modes.}\BibitemShut {Stop}%
\bibitem [{fus()}]{fussnote1}%
  \BibitemOpen
  \href@noop {} {}\bibinfo {note} {{T}he choice that each electron has the same
  momentum does not contradict our assumption of distinguishable particles
  since we only focus on the longitudinal momentum relevant for the scattering
  process. Indeed, taking into account other degrees of freedom, for example,
  the transverse extension of the electron bunch, makes the electrons
  distinguishable.}\BibitemShut {Stop}%
\end{thebibliography}
